\documentclass[conference]{IEEEtran}
\IEEEoverridecommandlockouts

\usepackage{cite}
\usepackage{amsmath,amssymb,amsfonts}
\usepackage{algorithmic}
\usepackage{graphicx}
\usepackage{subcaption}
\usepackage{textcomp}
\usepackage{xcolor}
\usepackage{multirow}
\usepackage{algorithm}
\usepackage{authblk}
\def\BibTeX{{\rm B\kern-.05em{\sc i\kern-.025em b}\kern-.08em
    T\kern-.1667em\lower.7ex\hbox{E}\kern-.125emX}}
\begin{document}

\title{From Agent Simulation to Social Simulator: A Comprehensive Review (Part 1)\\
}
\renewcommand\Affilfont{\small}
\author[1,2,3*]{Xiao Xue}
\author[4,5]{Deyu Zhou}
\author[6]{Ming Zhang}
\author[7]{Fei-Yue Wang}

\affil[1]{\textit{College of Intelligence and Computing, Tianjin University, Tianjin, China}}
\affil[2]{\textit{Tianjin Key Laboratory of Healthy Habitat and Smart Technology, Tianjin, China}}
\affil[3]{\textit{Laboratory of Computation and Analytics of Complex Management Systems, Tianjin University, Tianjin, China}}
\affil[4]{\textit{School of Software, Shandong University, Jinan, China}}
\affil[5]{\textit{Joint SDU-NTU Centre for Artificial Intelligence Research (C-FAIR), Shandong University, Jinan, China}}
\affil[6]{\textit{Faculty of Environment, Science and Economy, University of Exeter, Exeter, UK}}
\affil[7]{\textit{Institute of Automation Chinese Academy of Sciences, Beijing, China}}

\affil[ ]{ }
\affil[ ]{Email: jzxuexiao@tju.edu.cn, zhoudeyu@mail.sdu.edu.cn, mz427@exeter.ac.uk, feiyue@ieee.org}

\maketitle

\begingroup
\renewcommand\thefootnote{}\footnotetext{*Corresponding author: Xiao Xue, e-mail:jzxuexiao@tju.edu.cn.\\

**This work is an updated and extended version of our previous paper published in \textit{IEEE Transactions on Computational Social Systems} (DOI: 10.1109/TCSS.2021.3125287).
}
\endgroup

\begin{abstract}
This is the first part of the comprehensive review, focusing on the historical development of 
Agent-Based Modeling (ABM) and its classic cases. It begins by discussing the development history 
and design principles of Agent-Based Modeling (ABM), helping readers understand the significant 
challenges that traditional physical simulation methods face in the social domain. Then, it 
provides a detailed introduction to foundational models for simulating social systems, including 
individual models, environmental models, and rule-based models. Finally, it presents classic cases 
of social simulation, covering three types: thought experiments, mechanism exploration, and 
parallel optimization.
\end{abstract}


\section{What's ABM?}
\subsection{Development History of ABM}
In the past decade, Internet technology has effectively digitized the social, economic, political 
and cultural activities of billions of people, generating a massive amount of data. Based on these 
data, researchers view social systems as networks of interactions between their components and use 
complex network analysis to understand system characteristics, which has produced a series of 
landmark results. In 1998, DJ Watts, a doctoral student at Cornell University, and his mentor 
published a paper titled “Collective dynamics of ‘small-world’ networks” in Nature \cite{Watts1998}; in 1999, 
AL Barabási, a professor at the University of Notre Dame, and his doctoral student R. Albert 
published a paper titled “Emergence of Scaling in random networks” in Science \cite{Barabasi1999}. The two papers 
revealed the small-world characteristics and scale-free properties of complex networks, 
respectively, marking the beginning of a new era in complex network research.

However, the data studied by complex network analysis (such as search and social media data) are 
usually not generated for specific research questions, and are often noisier, less structured, and 
less systematically “designed” than traditional social science data (such as surveys and laboratory 
experiments). In this context, simple data analysis is in a dilemma in terms of 
methodology: (1) Data representativeness: Big data is also a form of sampling, and its 
representativeness is difficult to evaluate. People who use the Internet and various social 
platforms are not randomly distributed in the total population. For example, the elderly use the 
Internet less. When representativeness cannot be guaranteed, the huge sample size of big data 
cannot increase the reliability of research conclusions. (2) Data authenticity: The individual or 
group behavior patterns obtained from big data may not be consistent with those in real life. For 
example, some people are very active on the Internet, but they are taciturn and unsociable in real 
life. In addition, big data only records some aspects of people's lives, not all aspects. (3) Data 
reliability: In many cases, data are records left when we use a certain product (such as search 
engines, Toutiao). It should be noted that when using these products, our behavior patterns may be 
guided by recommendation algorithms and undergo some changes, which will cause the corresponding 
data to be “distorted”.

\begin{figure}[htbp]
\centering
\includegraphics[width=\linewidth]{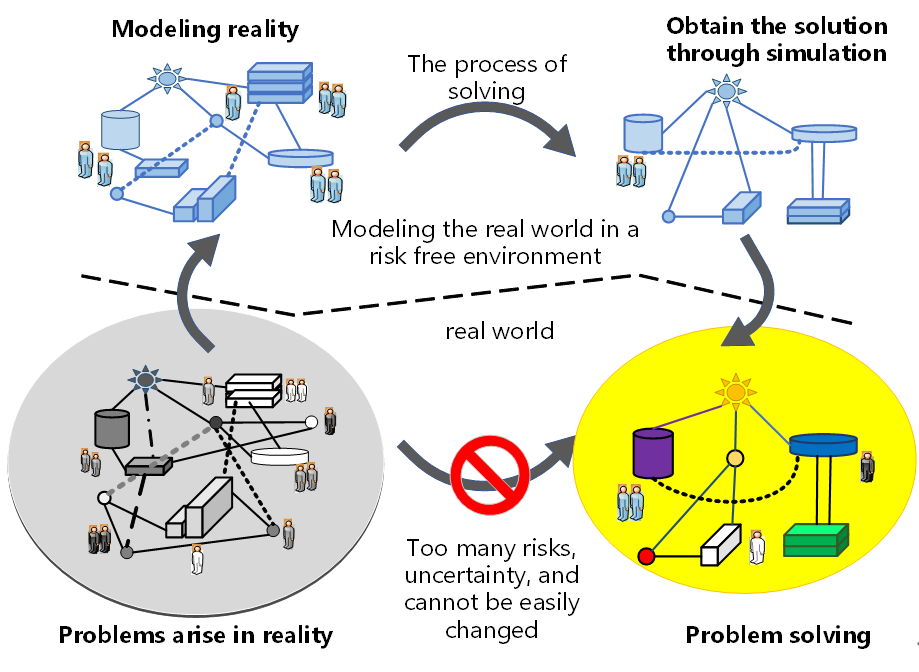}
\caption{The core idea of social simulation.}
\end{figure}

Field experiments on social systems can solve the above problems, but they often face huge 
challenges in practice. The reasons can be attributed to the following aspects: (1) Economically, 
due to the scale and cost factors of social systems, the cost of experiments is too high to be 
affordable; (2) Legally, issues such as national defense, military preparedness, and social 
security are protected by legislation, so it is impossible to experiment on the systems under 
research or rebuild these systems for experiments; (3) Morally, social systems often involve a 
large number of people. Experimenting on these systems may have an impact on people's normal lives 
and even endanger people's lives and property, so such experiments are morally unacceptable. In 
this context, social modeling and simulation methods have gradually developed into another 
mainstream method in computational social science research. As shown in \textbf{Figure 1}, social 
simulation uses computers as “artificial laboratories” to “cultivate” macro phenomena that may 
appear in actual systems, explore the laws behind them, and provide a feasible way to analyze the 
behavior of complex systems and evaluate the effects of interventions. There are two types of 
social modeling and simulation methods: 

\textbf{Top-down modeling method:} The typical representative is the system dynamics method, which uses 
mathematical language to express the characteristics, states, relationship and process of a 
specific object or problem, and forms explanations, makes judgments and predictions through 
deduction, calculation and analysis. This type of method is highly abstract and logically 
rigorous, but when faced with nonlinear systems, it may be the case that the system can be 
represented but not solved. At the same time, the model results are very sensitive to the model 
assumptions, resulting in the risk of “a slight error leading to a huge mistake”.

\textbf{Bottom-up modeling method:} The typical representative is Agent-Based Modeling (ABM), 
which abstracts and simplifies the behavior of the subjects in the system into agents, and 
produces interesting macro results through simple interactions between them. It is considered to 
be the third scientific method besides classical deductive reasoning and inductive reasoning \cite{Axelrod1981}. 
Compared with the top-down modeling method, the advantage of ABM is mainly reflected in the 
inclusion of the framework of emergence. The basic elements of the model need to be flexible 
enough, including the heterogeneity and adaptability of the subjects, scalability of quantity, 
etc., so that new and unexpected features will naturally appear in the model.

In the 1990s, the Santa Fe Institute first began to use computer simulation to address economic 
system problems, pioneering social simulation research. For the first time, Epstein and Axtell's 
“Growing Artificial Society” systematically used ABM as a modeling tool for social sciences, that 
is, to build an artificial social laboratory through the KISS principle (Keep It Simple and 
Stupid)\cite{Epstein1996}. ABM is related to the concept of multi-agent systems but differs from it. Its goal is 
to provide explanatory insights into the collective behavior of agents that follow simple rules in 
the system, rather than to design agents or solve specific practical or engineering problems. ABM 
is mainly used to achieve the following purposes: (1) Existential experiments: to observe whether 
a certain set of rules can lead to the emergence of a certain complex behavior; (2) Understanding 
experiments: to try to simulate the system from the bottom up to understand the operation process 
of complex systems , especially those social complex systems that cannot be actually experimented 
on; (3) Deductive experiments: to be used to solve large-scale equations for tasks such as 
“predicting world population dynamics or global epidemic spread”. Traditional mathematical 
solutions are often powerless for such problems.

The ABM toolbox is applicable to research involving complex interactive processes in all 
disciplines, ranging from urban traffic simulation to humanitarian aid. The technologies and 
applications involved in ABM are very complex, and its connotation and scope are constantly 
changing over time. \textbf{Figure 2} shows the important topics in ABM research, which can be divided 
into three periods based on the timeline of events: the first stage (1970-1995) is the formation 
stage of the basic concept of social simulation, including “artificial life” and “artificial 
society”; the second stage (1995-2004) sees the emergence of artificial society models in various 
fields (such as sociology, business and marketing, land use, finance, etc.), and the problems 
existing in ABM itself have also begun to attract attention; the key event of the third stage 
(2004- present) is the emergence of the ACP method, which marks the beginning of the shift of 
ABM from “toy models for verifying hypotheses” to "how to empower reality". With the emergence 
of large language model technology, the field of agent-based simulation is entering a stage of 
rapid development. \textbf{Table 1} shows the landmark events in this process.

\begin{figure*}[htbp]
\centering
\includegraphics[width=15cm]{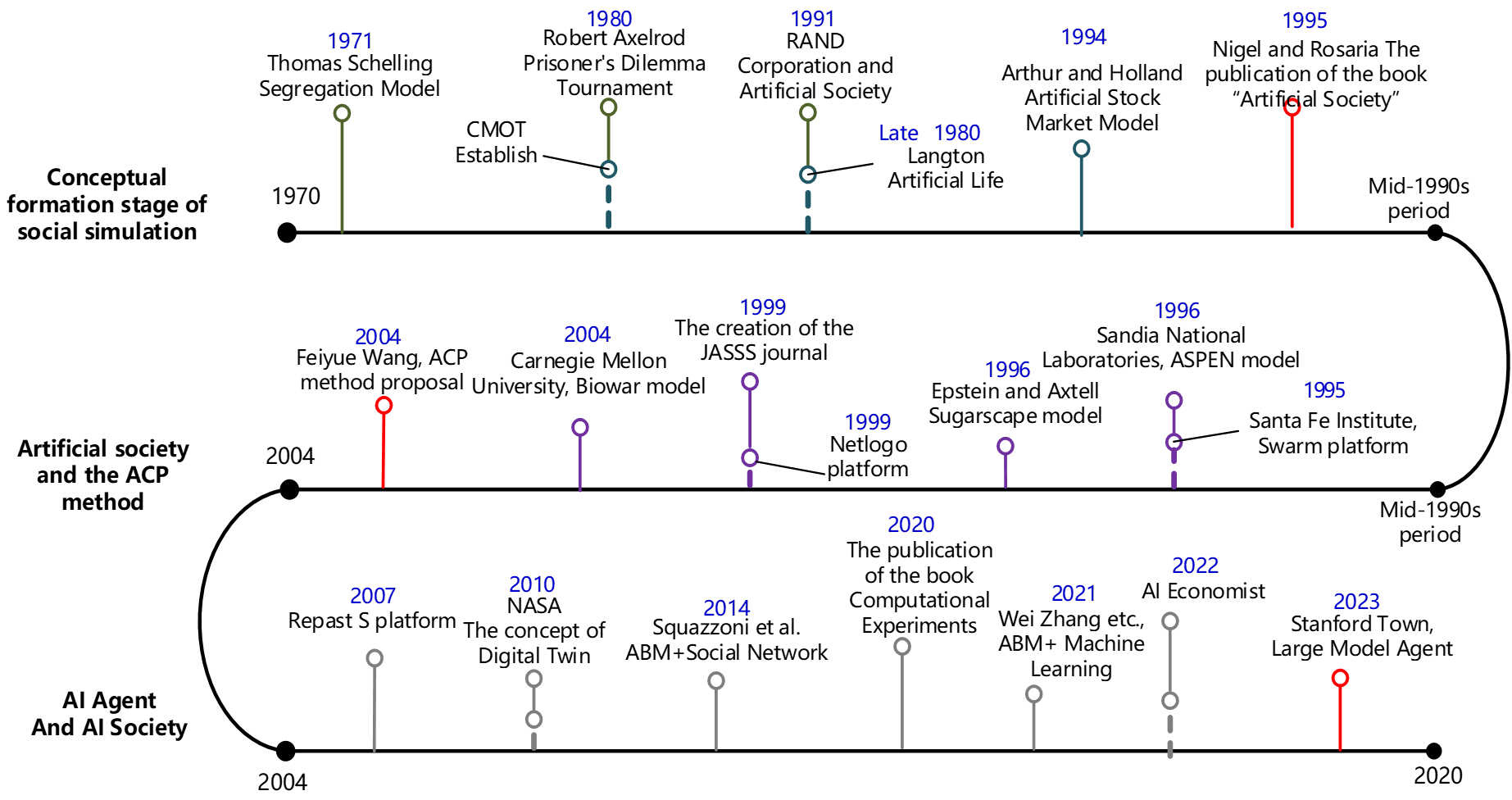}
\caption{The development of ABM.}
\end{figure*}

\begin{table*}[!t]
\centering
\caption{ABM development history.}
\label{tab:1}
\begin{tabular}{|p{0.5cm}|p{2cm}|p{14cm}|}
\hline
Years & Figure & Representative Work \\
\hline
1970 & Conway & Wrote the “Game of Life” program, opening the prelude to artificial life research \cite{Gardner1970}. \\
\hline
1971 & Thomas Schelling & The classic Segregation model \cite{Schelling1971} was proposed, which embodies the basic concept of the agent model, that is, autonomous agents interacting in a shared environment can eventually observe aggregation and emergence phenomena. \\
\hline
Early 1980 & Robert Axelrod & Focused on modeling political systems, covering phenomena such as racism and cultural communication. Hosted the Prisoner's Diemma strategy competition \cite{Axelrod1981}. \\
\hline
Mid 1980 & CMOT & CMOT(Computational and Mathematical Organization Theory) was founded, and the field grew as a special interest group of the Management Science Institute and its sister society, the Operations Research Society. \\
\hline
Late 1980 & Christopher Langton & Proposed the theory of artificial life and, together with other researchers, used computers to construct model systems with behavioral characteristics of natural life systems \cite{langtonArtificialLifeII1994}, such as self-reproducing cellular automata, Boids model, ant colony models, Tierra models, and “Amoeba World”. \\
\hline
1991 & RAND Corporation, USA & For the first time, the concept of “artificial society” was formally proposed, which is to use agent technology to build a social laboratory (subjects, environments and rules) in a computer to conduct experimental evaluations on different policies to ensure the effectiveness of the policies \cite{Builder1991}. \\
\hline
1994 & Arthur and Holland & Based on the theory of complex adaptive systems, a research framework for artificial stock markets was proposed \cite{arthur2018asset}, emphasizing that financial markets are complex dynamic systems formed by the interaction of heterogeneous entities with learning capabilities. \\
\hline
1995 & Nigel Gilbert and Rosaria Conte & The book “Artificial Societies: The Computer Simulation of Social Life” was published \cite{Gilbert1995}, and the concept of artificial society was formally proposed and became a relatively independent field of social science. \\
\hline
Mid 1990s & Santa Fe, USA & Released the Swarm platform, which was initially developed by Chris Langton, Roger Burkhart, Nelson Minar and others, and subsequently many people around the world participated in its open source follow-up development. \\
\hline
1996 & Sandia National Laboratories, USA & The ASPEN model \cite{Basu1998} is an agent-based microeconomic simulation model designed to simulate and study the U.S. economy in detail and is widely used in the field of economic analysis. \\
\hline
1996 & Epstein and Axtell, University of Chicago & Developed “Sugarscape” to simulate and explore social phenomena such as seasonal migration, sexual reproduction, genetic inheritance, and market formation \cite{Epstein1996}. \\
\hline
Late 1990s & CASOS & CMOT was renamed Computational Analysis of Social and Organizational Systems (CASOS). The North American Association for Computational Social and Organizational Sciences (NAACSOS), the European Social Simulation Association (ESSA), and the Asia-Pacific Social Simulation Association (PAAA) were also established. \\
\hline
1999 & Nigel Gilbert & The book “Simulation for the social scientist” was published and the journal “The Journal of Artificial Societies and Social Simulation”(JASSS) was founded, marking the gradual maturity of the field of sociological simulation. \\
\hline
Early 2000 & Uri Wilensky, Northwestern University, USA & Netlogo was developed with the characteristics of easy operation and good visualization. It is very suitable for beginners to model and simulate multi-agent systems and is widely used in education, scientific research and other fields. \\
\hline
2004 & Carnegie Mellon University (CMU), USA, Kathleen Carley & The Biowar model was developed to combine computational models such as social networks, transmission vectors, disease models, demographic models, spatial models, wind diffusion models, and diagnostic models into a single integrated system designed to simulate the impact of a bioterrorist attack on a city \cite{Carley2006}. \\
\hline
2004 & Institute of automation, Chinese Academy of Sciences, Wang Fei-Yue & A set of ACP methods for studying complex systems, namely “artificial system + computational experiment + parallel execution”, was proposed, which uses computational experiments to provide “reference”, “prediction” and “guidance” for possible situations of actual system operation \cite{Wang2004ParallelSM, Wang2007}. \\
\hline
2007 & Michael J. North et al. & The multi-agent simulation tool Repast was developed \cite{North2007}; Repast Simphony version 2.0 was released in 2012; it is still being updated and the latest version was released in 2024. \\
\hline
2010 & NASA & The concept of digital twins was proposed, emphasizing the establishment of a virtual-real two-way dynamic feedback mechanism, which can simulate, analyze and optimize physical entities. This technology can greatly enhance the fidelity of environmental modeling. \\
\hline
2014 & Squazzoni et al. & Emphasized the combination of ABM and network science \cite{Squazzoni2014}, mapping the interactions between agents into nodes and links in the network can supplement the network structure sampling bias commonly found in social empirical methods. \\
\hline
2020 & Xue Xiao & Published “Computational Experiment method for Complex Systems: Principles, Models and Cases”\cite{Xue2020}, the first book to systematically review computational experiment method. \\
\hline
2021 & Wei Zhang et al. & Studied how to combine different Machine Learning (ML) techniques with ABM to model complex, large-scale systems from the perspective of improving model accuracy or robustness and providing better decision-making strategies \cite{Zhang2021}. \\
\hline
2022 & S Zheng et al. & Proposed a new study called “The AI Economist” was proposed, which uses economic simulation methods to discover tax strategies that can effectively find a balance between economic equality and productivity \cite{Zheng2022}. \\
\hline
2023 & Stanford University, USA, Joon Sung Park, etc. & A generative agent architecture based on a large language model was proposed, and a virtual sandbox environment “Stanford Smallville” was designed \cite{Park2023}. 25 agents with unique backgrounds run in it, and the complete experience of the agents can be stored through natural language, with features such as task decomposition, memory, and reflection. \\
\hline
2024 & Xue Xiao \& Wang Fei-Yue & n response to the challenges faced by Agent-Based Modeling (ABM) in the realm of intelligent decision-making, Xue Xiao and Wang Fei-Yue conducted in-depth research on the methodological framework \cite{xiao2023putational, xue2023computational} and key steps \cite{xue2024computational, xue2024computational4, xue2023computational5, lu2021computational, xue2021computational7} of computational experiments, and provided several case studies \cite{xue2021soa, xue2019analysis, xue2020value, xue2016computational, xue2016computational12, xue2019value}. \\
\hline
\end{tabular}
\end{table*}

\subsection{Design Principles of ABM}
All social simulations are designed to solve one or more research problems defined based on a 
reference system. Research problems are the main engine of scientific exploration and largely 
determine the degree of consistency between social simulations and the real world. Generally 
speaking, the representations of social simulations is relatively simple, while the operation 
of actual systems is relatively complex. The contradiction between the two is mainly manifested 
in two aspects: (1) If the level of abstraction of the simulation system is too high, it will 
be difficult to reflect the operational laws of the real world; (2) If the level of 
abstraction of the simulation system is too low, the complexity of model building will be 
very high, and a series of problems such as lack of data, insufficient resources, and 
inadequate knowledge system will be encountered. Therefore, it is very important to achieve 
an appropriate mapping between the real world and the simulation system. In the design of 
simulation models, the model design principles should be balanced with the actual situation, 
as shown in \textbf{Figure 3}.

\begin{figure}[htbp]
\centering
\includegraphics[width=\linewidth]{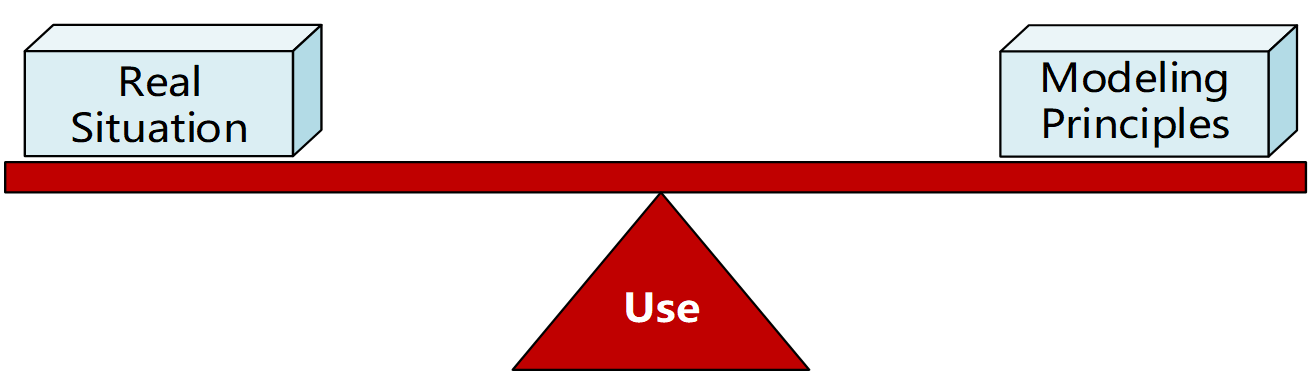}
\caption{The balance between model design principles and actual conditions.}
\end{figure}

Model design is a balance between reality and abstraction at a certain level, so it is not 
expected that the model will perform well at different levels. For example, a highly realistic 
model may have great policy value but little or no theoretical value; conversely, a highly 
abstract model may provide profound scientific insights but fail to provide directly 
applicable results in terms of policy contribution. Two modeling principles are given below, 
depending on the degree of closeness to a given reference system:

\paragraph{- KISS principle (abstract)}
The KISS (Keep It Simple and Stupid) principle is mainly applicable to theory building, where 
the model is small in scale and has few parameters, and the goal is to find the minimum 
counter-intuitive conditions for a certain macro-emergent phenomenon \cite{Flache2021}. The KISS-oriented 
ABM is regarded as a “thinking tool” or “intuitive engine”, and the balance between the model 
design principles and the real situation is shown in \textbf{Figure 4}. This model is sometimes 
called a toy model, which is similar to the reference system only in a few qualitative aspects 
and does not attempt to reproduce any quantitative characteristics, but it can provide some 
applicable insights in the problem domain. Supporters of the KISS method agree with the 
following view: “If the goal of ABM is to enrich our understanding of the emergence process of 
a phenomenon…, then the simplicity of the assumptions is important, and the reproduction of 
the real details in a specific environment is not so important” \cite{Axelrod}.

\begin{figure}[htbp]
\centering
\includegraphics[width=\linewidth]{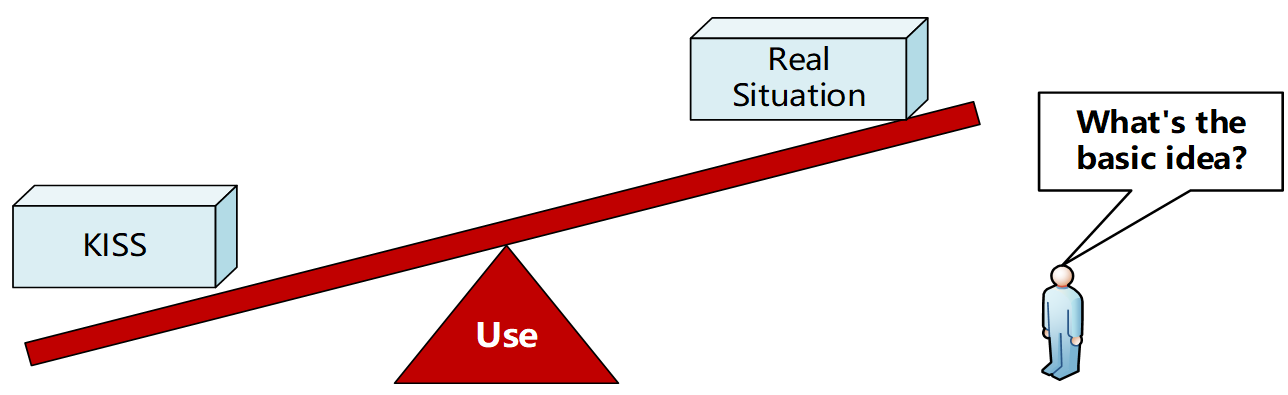}
\caption{The balance between the design principles of the parsimony method model and the actual situation.}
\end{figure}

Many early social simulations belong to this category, providing a unique way to understand 
the basic rules of human and social operations. The widely praised racial segregation 
model \cite{Schelling1971} assumes an ideal two-dimensional space in which “stars” and “zeros” decide whether 
to change their positions based on the characteristics of their nearest neighbors. Schelling 
did not use any sociological or psychological theory to prove his microscopic assumptions, nor 
did he use any empirical data to set the preferences of agents, neighborhood ranges, and group 
sizes. Instead, Schelling made (weak) structural analogies between the fictional mechanism and 
the segregation mechanism in reality based on intuition and common sense. The results of the 
model are not specific to racial segregation data in a specific geographic area, but can 
actually cover any phenomenon in which two groups have some tendency to separate. Here, the 
realism of the model is secondary, and the heuristic value of the model is what is really 
important.

Subsequently, the Sugarscape model designed by Epstein and Axtell can be seen as a typical 
case of the KISS principle. The assumptions about the agent rules in the model use some 
existing sociological/psychological theories, but they are weak; data are also not used to 
assess the extent to which the simulation results reproduce real-world regularity. For this 
reason, this type of ABM only has heuristic value, but it is not clear to what extent the 
model reflects the real world. Epstein and Axtell's book “Growing Artificial Society” has had 
a considerable impact on beginners to ABM \cite{Epstein1996}. Since then, many studies have been dominated 
by the Sugarscape style, including the evolution of cooperation \cite{Axelrod1981}, trust and 
reputation \cite{Pinyol2013}, and the emergence of norms \cite{Axtell2007}. For a long time, the field was dominated by 
abstract ABMs, far more than data-driven ABMs.

\paragraph{- KIDS principle (realistic)}
Although historically, the KISS principle has had a clear impact on the field, the knowledge 
generated by such “unfounded” ABM, which lacks credibility, has faced a series of skepticism. 
Modelers, accustomed to the KISS principle, are beginning to seek ABMs that are more closely 
connected to the “real world”. In this context, the KIDS (Keep It Descriptive and Stupid) 
modeling principle began to emerge. This type of model is oriented to solving application 
problems, as close to reality as possible. The model is large-scale, has many parameters, 
requires testing of many modules, and its experimental output is most consistent with 
empirical data. \textbf{Figure 5} shows the balance between the design principles of the model and 
the real situation. This type of model is widely used in commercial and government agencies. 
For example, public policy involves areas of high uncertainty (such as social networks and 
human behavior), and ABM can be used to analyze the effects of policies (such as economic 
stimulus, laws and regulations, etc.) to improve the scientific basis of public policy making.

\begin{figure}[htbp]
\centering
\includegraphics[width=\linewidth]{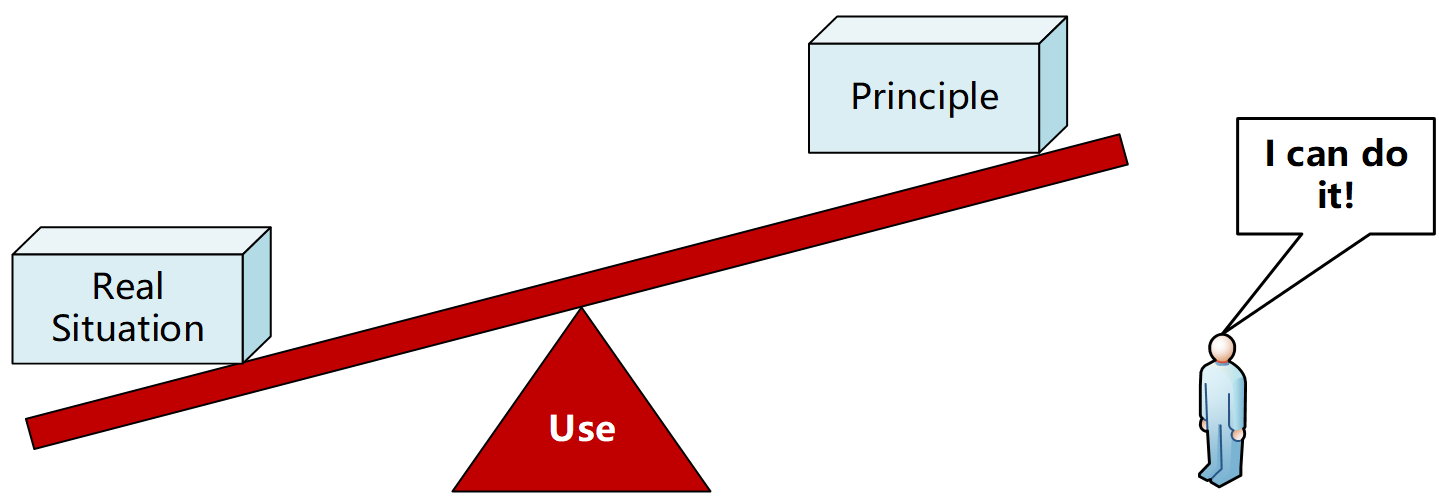}
\caption{The balanced relationship between realistic method model design principles and real situations.}
\end{figure}

Proponents of KIDS believe that ABM has very high flexibility in mechanism design, and model 
simplification is only a possible result of analysis, not a starting point for analysis \cite{Moss2005a}. 
From the KIDS perspective, a model that is rich in scope should be first established to allow 
for the description of existing empirical evidence, and then simplified; the simplification 
process cannot be contrary to the empirical data of the phenomenon being studied and will not 
degrade the model's performance in interpreting the phenomenon. The goal is to design a “high 
fidelity” ABM \cite{DeMarchi2014}, that is, both the micro assumptions of the model and the macro objectives 
that are to be reproduced have “high-dimensional authenticity” characteristics \cite{Bruch2015}.

The modeling method of KIDS has begun to deeply influence contemporary ABM research, based on 
real-world datasets, initializing the probability distribution of ABM core variables, 
formalizing the network structure between agents, and establishing agent behavior rules. 
Related cases include the differential fusion of innovation \cite{Roth2010}, reputation dynamics \cite{Boero2010}, 
and game theory of cooperative behavior \cite{Wunder2013}. In economics, ABM has also begun to combine with 
experimental research on economic decision-making \cite{Duffy2006} and empirical verification \cite{Windrum2007}. 
Therefore, while “calibration and verification of models” remains a research frontier in the 
field, ABM is increasingly moving from exploratory models to rigorous empirical verification.

\begin{table*}[!t]
\caption{Application classification of ABM}
\centering
\begin{tabular}{|p{2cm}|p{2cm}|p{12.5cm}|}
\hline
\textbf{Case category} & \textbf{Year and people} & \textbf{Specific case content} \\
\hline

\multirow{4}{=}{\textbf{Theoretical support}} 
& 1965, Torsten Hägerstrand & ABM was used to explain the agglomeration patterns that emerged in terms of innovation on farms in two regions of Sweden \cite{Hagerstrand1965}. In the two-dimensional space, the distribution of farms in the relevant regions of Sweden is replicated, and the contact probability matrix is set up based on local telephone traffic and migration statistics. \\
\cline{2-3}
& 2001, 2003, Janssen and Jager & ABM was used to study the diffusion of products in the market \cite{Janssen2003}. Specific psychological theories of preference change are used to design the behavior rules of the agents. All the parameters and variables of ABM ultimately come from theoretical probability distributions, which have a certain degree of truth in terms of network composition. \\
\cline{2-3}
& 2007, Billari et al. & ABM was used to explain the similarity in the age distribution of first marriage across countries \cite{Billari2007}. A cognitive heuristic agent was designed to simulate mate selection, and the relative explanatory power of each model variant was evaluated by comparing the simulation distribution with the empirical distribution. \\
\cline{2-3}
& 2013, Mäs and Flache & ABM was used to study polarization of opinion in the community \cite{Mas2013}. The micro hypothesis comes from the theory and empirical evidence of social psychology about attitudinal changes and memory processes; some model parameters were calibrated by experimental data. Finally, the simulated opinion trend is compared with the experimental results. \\
\hline

\multirow{4}{=}{\textbf{Parameter calibration}} 
& 2008, Dugundji and Gulyas & Individual commuting patterns in the city of Amsterdam were studied using ABM \cite{Dugundji2008}. The agent's choice of transportation mode is based on the survey data, and a function of socio-demographic characteristics (such as gender, income, age, education, and place of residence) is constructed, which can predict the proportion of agents choosing different transportation modes. The Agents' behavior does not refer to sociological or psychological theories. \\
\cline{2-3}
& 2011, DiMaggio and Garip & The adoption of ABM explains the inequality in the rate of adoption of new technologies among people with different educational backgrounds \cite{DiMaggio2011}. The size of the agent population, as well as individual attributes, such as income, education, ethnicity, and network size, are all based on survey data from the United States. \\
\cline{2-3}
& 2014, Fountain and Stovel & ABM was used to study the role of network structure and referral in career instability \cite{Fountain2014}. The distribution of employee skills and company size is calibrated based on U.S. empirical data. \\
\cline{2-3}
& 2011, Bruch and Mare & Employed ABM to explore levels of apartheid \cite{Mare2011} A function for estimating actors' racial preferences based on survey data from American films was used. \\
\hline

\multirow{2}{=}{\textbf{Verification of results}}
& 1999, Lux and Marchesi & ABM was used to explain the recurring statistical characteristics of financial markets \cite{Lux1999}, such as the distribution of unconditional returns, the long tail or the persistence of price fluctuations. Trading entities switch from one group to another based on the comparison of their respective profits, opinion indices and price trends. \\
\cline{2-3}
& 1997, Arthur et al. & ABM was used to explain price fluctuations in the stock market \cite{arthur2018asset}. Agents modify trading strategies by performing random mutations and cross-operations on trading rules. The ability of ABM to reproduce the statistical characteristics of real stock markets, as well as the presence of behavioral heterogeneity, is used to compensate for the lack of realism of micro-level assumptions and the lack of attention to input-level calibration. \\
\hline

\textbf{Parameter calibration + Verification of results} & 2011, Ajelli et al. & An epidemic on a national scale was simulated using ABM \cite{Poletti2011}. Based on various census data, the agents are assigned socio-demographic characteristics, geographical location, and probability of action. The geographical space at the city level as well as the physical location of the workplace are also reflected. This model is used to predict the evolution of similar diseases at multiple scales (i.e., national, regional, and urban) and in different populations. \\
\hline
& 2011, Frias Martinez et al. & ABM was used to assess the impact of government epidemic prevention interventions \cite{Frias-Martinez2011}. The model was calibrated and tested based on data from the 2009 H1N1 influenza outbreak in Mexico. Used cell phone recordings to characterize the true network and flow of connections between agents, allowing for finer parameter calibration. These models often do not refer to specific sociological or psychological theories, leaving “theoretical realism” at the micro level close to zero. \\
\hline

\end{tabular}
\label{tab:abm_cases}
\end{table*}

\subsection{Causal challenges for ABM}
In general, the ABM model is very simple and (relatively) easy to fully examine and understand 
its internal functioning. Thus, the knowledge they produce can be explained from a generative 
perspective (vertical causality on the right side of \textbf{Figure 6}) in terms of causality, i.e., 
which dynamic chain of events relates the model's micro-hypotheses to a given set of 
macroscopic outcomes. In fact, “intelligibility” is thought to help enhance the “credibility” 
of abstract models \cite{Sugden2000}, and Schelling’s model of apartheid is one such model. If we look at 
it from the perspective of causal dependence (horizontal causality on the left side of 
\textbf{Figure 6}), this knowledge can also be explained causally. By manipulating ABM parameters 
(i.e., sensitivity analysis) and changing the model internals (i.e., robustness analysis), it 
is indeed possible to determine that a certain parameter of the model is responsible for 
changes in the macro pattern.

\begin{figure*}[htbp]
\centering
\includegraphics[width=15cm]{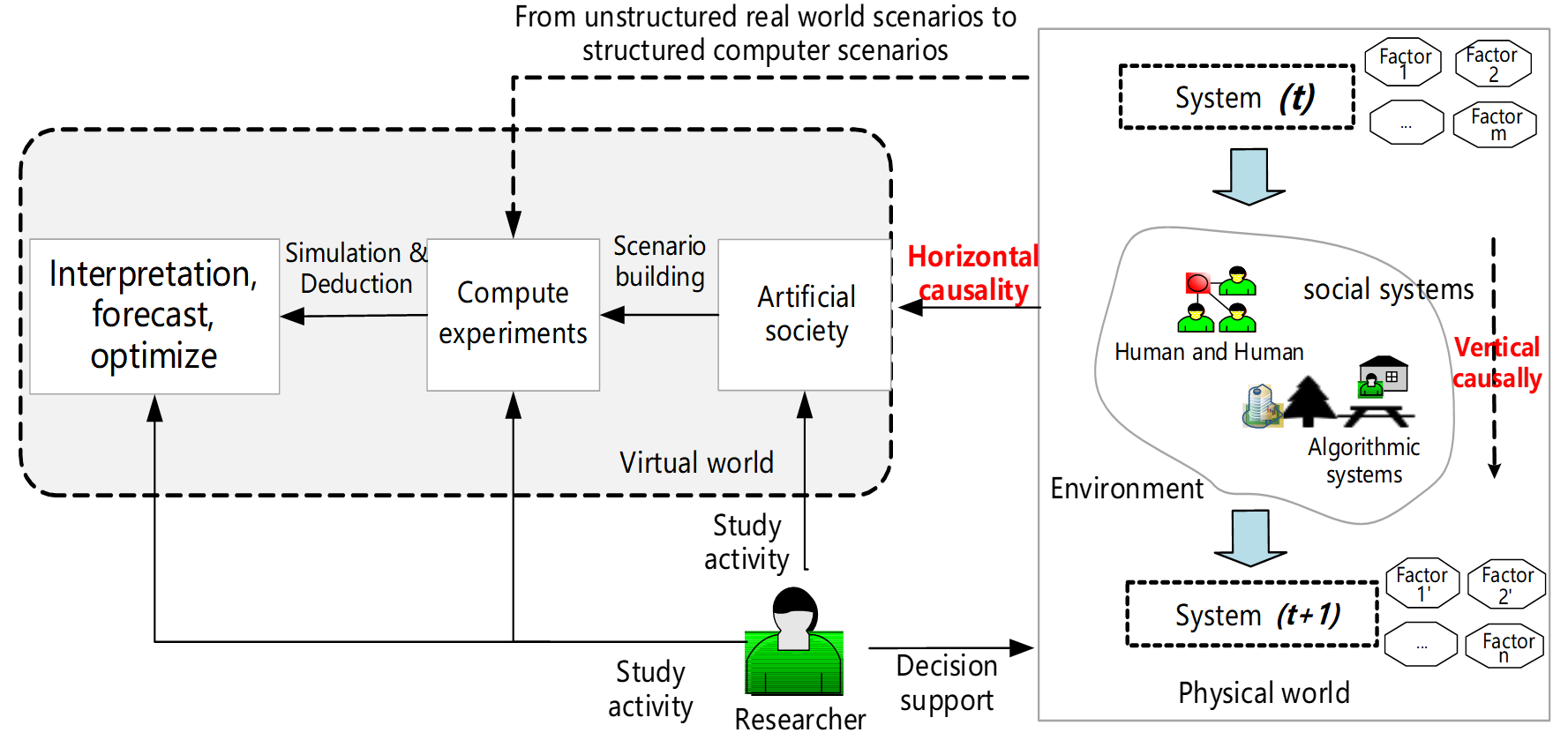}
\caption{Schematic diagram of vertical and horizontal causality of ABM.}
\end{figure*}

However, the explainable causal knowledge generated by the abstract model is purely internal 
knowledge of the ABM instantiation system. In terms of inputs, it is not possible to argue 
that the micro-mechanisms of ABM mimic any aspect of social reality, as there are no specific 
theories or data to support hypotheses at the micro level. In terms of output, it is unclear 
what ABM is actually replicating because the model's macroscopic numerical results do not 
coincide with any clear empirical laws. Therefore, neither from the perspective of causal 
dependence nor from that of generation, do we have access to causal knowledge of any 
micro-level mechanism or the dynamic processes that arise from it. A highly abstract ABM 
obviously cannot be used directly for causal reasoning.

With the maturity of ABM, how to conduct causal inference based on simulation technology to 
provide decision support for real-world scenarios has become the focus of attention. Although 
many researchers (see \cite{Axtell2002, Cederman2005, Tesfatsion2006}) all recognize ABM as a methodological tool based on mechanism 
explanations, there are doubts about the extent to which ABM can demonstrate that the 
hypothesized mechanisms can work in the real world. The literature \cite{DiezRoux2015} describes it this way:

\begin{quote}
``(...) \textit{There is a fundamental difference between causal inference based on 
observation (such as traditional epidemiology) and causal inference based on simulation 
modeling. Traditional epidemiological tools extract reasonable conclusions from incomplete 
real-world observations. (…) However, when we use ABM, we create a simulated world (based on 
prior knowledge or intuition), and then explore hypotheses about causes in this simulated 
world. These hypotheses are implemented through coding in the model. In ABM, we cannot 
directly determine whether X causes Y in the real world (because the virtual world is created 
by ourselves); In the model, we can only explore the possible impact of changing the X on 
the Y level. In the real world, we have facts and try to infer counterfactual conditions (what 
would we observe if handled differently). In the simulated world, everything is counterfactual 
because the world and all possible scenarios are artificially created by the researchers.}'' 
\end{quote}

Diez Roux’s statement proves a broad belief that observational and experiment method are 
fundamentally different from ABM: the former deal with “facts” while the latter consists 
entirely of “hypotheses”; the former allows us to make causal connections in the real world, 
while the latter is considered by many to be impossible. Thus, the “observational approach” 
has an inherent superiority in causal inference, allowing for a unified counterfactual 
explanation within a latent causal framework \cite{Imbens2015, Morgan2014}. These methods rely on “partial and 
incomplete (often chaotic) observations”, which are regarded as self-evident. However, it is 
important to note that observation methods are forced to rely on assumptions precisely because 
of data limitations. ABM, on the other hand, is seen as relying solely on “prior knowledge or 
intuition”, which is considered "one of the most troubling problems" of the approach. How to 
systematically combine ABM with observational and experimental methods to maintain a balance 
between empirical data and theoretical hypotheses has become a problem for researchers.

For traditional observational experiments, the difficulty lies in how to construct 
counterfactual scenarios. But for ABM, the parameters of the simulation environment can be 
systematically controlled, and building counterfactual scenarios is no longer a challenge. 
Instead, how to construct a benchmark scenario that reflects reality becomes a new challenge, 
including whether the simulator truly reflects the data generation process and whether the 
final simulation results are reliable. In order to achieve this goal, ABM can also be anchored 
to data, and through parameter calibration and empirical verification, the theoretical 
realism and the realism of “input” and “output” can be achieved, respectively. This makes ABM 
a kind of “simulator” that allows the modeler to use ABM’s knowledge across different levels 
to make inferences about the mechanics of the world outside the model. Due to data 
availability, ABM is often unable to perform a full empirical calibration, and reliability 
tools can only be used to quantify the degree to which ABM relies on modeling assumptions. 
Therefore, if an ABM achieves maximum empirical calibration and validation, and is checked 
for system reliability, it is well placed to support the relevant causal inferences (i.e., 
connections between different levels in the real world). \textbf{Table 3} shows the history of 
researchers in this field.

\begin{table*}[!t]
\centering
\caption{ABM works with causal inference.}
\label{tab:3}
\begin{tabular}{|p{2.5cm}|p{11cm}|p{3cm}|}
\hline
\textbf{Scholars} & \textbf{Main points} & \textbf{Evolution of perspectives} \\
\hline
Grüne-Yanoff (2009a) & ABM cannot provide a “complete” causal explanation because it is impossible to empirically verify all components of the model design mechanism; ABM cannot support “possible causal history” because there is a lack of an internal standard to select from all possible options \cite{Grune-Yanoff2009}. & Studied the Anasazi model, based on which they denied the usefulness of ABM for causal inference. \\
\hline
Elsenbroich (2012) & ABM can guide or complement causal inference methods that rely on empirical evidence, but it cannot independently determine the actual causal story, and complete understanding of micro-phenomena is impossible for causal explanations \cite{Elsenbroich2012}. & This is not a problem for ABM, but for social sciences as a whole. \\
\hline
Morgan\&Winship (2015) & Simulation is at best a tool for “theory building” and its utility is limited due to the lack of transparency. Although A BM can be considered as an analytical tool that bridges levels of analysis, it still lacks a solid methodology \cite{Morgan2014}. & ABM lacks a reliable methodology. \\
\hline
Macy\&Sato (2008) & Implicitly, a division of labor is proposed: ABM is a tool for theoretical exploration, while experimental and statistical methods are more suitable for observational data to support causal inference \cite{Macy2008}. ABM can generate hypotheses for testing, but cannot bear the burden of proof. & ABM and experimental statistics should be combined. \\
\hline
Bruch\&Atwell (2015) & ABM can communicate with empirical data in various ways, allowing it to identify the realistic mechanisms behind the dependencies between variables \cite{Bruch2015}. & ABM should absorb reality. \\
\hline
Casini (2014) & Whether a model is “credible” depends on the following three conditions \cite{Casini2014}: “(i) the plausibility of the theoretical rationale, psychological assumptions, and functional analogies; (ii) the robustness of the results to changes in initial conditions and parameter values; and (iii) the robustness to changes in modeling assumptions.” Robustness analysis is particularly important, demonstrating that the model in question captures the true features of the actual causal story, rather than accidental or artificial features. & The path to combining ABM with reality. \\
\hline
Ylikoski\&Aydinonat (2014) & Systematic variations of initial model assumptions (i.e., robustness analysis) lead to “model clustering” that, on the one hand, helps to propose “core” feature hypotheses about actual causation and, on the other hand, helps to propose alternative explanations, positioning a given causal story within a set of possible alternative causal stories for empirical testing \cite{Ylikoski2014}. & The specifics of ABM robustness analysis. \\
\hline
Törnberg (2019) & Because of the contingency of the social domain, it is impossible to know whether the simulated mechanisms are reflected in reality. The idea of carefully validating or calibrating the model by matching the regularities produced in the model with empirical regularities is “misleading” and “stems from confusing abstract categories with concrete phenomena”\cite{Tornberg2019}. & ABM should only be viewed as “an aid to enhanced intuition”. \\
\hline
Manzo (2022) & Experimental and observational methods face the same demanding challenges as ABM when used to establish “horizontal” causal claims or counterfactual dependency claims. Experimental observational methods can only make credible causal inferences about whether changes in an independent variable produce differences in the dependent variable through a mixture of data and argument\cite{Manzo2022}. & ABM and experimental observation methods complement each other. \\
\hline
\end{tabular}
\end{table*}

\section{How to Realize ABM?}

Social system simulation, similar to physical simulation, also involves stages such as the 
conceptual model, domain model, and computational model, with interdependent and constrained 
relationships among them. At the conceptual layer, agent models with human-like characteristics 
are constructed, including functional modules such as perception, cognition, decision-making, 
and behavior\cite{xue2018social, zhou2022sle2, zhou2024hierarchical}. The domain layer model is closely related to the conceptual layer and 
requires the development of a series of domain-specific models (such as geographic models, 
infrastructure, population models, intervention measures, and social relationships) to build 
an artificial society foundation consistent with the real world. At the numerical computation 
layer, agent strategy iteration equations based on neighborhood and memory information are 
constructed, reflecting a dynamic process of agent evolution and contextual updates. As shown 
in \textbf{Figure 7}, the digital thread serves as a bridge connecting different models, revealing 
the evolution of the system, historical configuration changes, and specific state transitions 
throughout the experimental cycle. In the following, we provide a detailed description of the 
three core modeling units of ABM, including the agent model, the environment model, and the 
rule model.

\begin{figure*}[htbp]
\centering
\includegraphics[width=15cm]{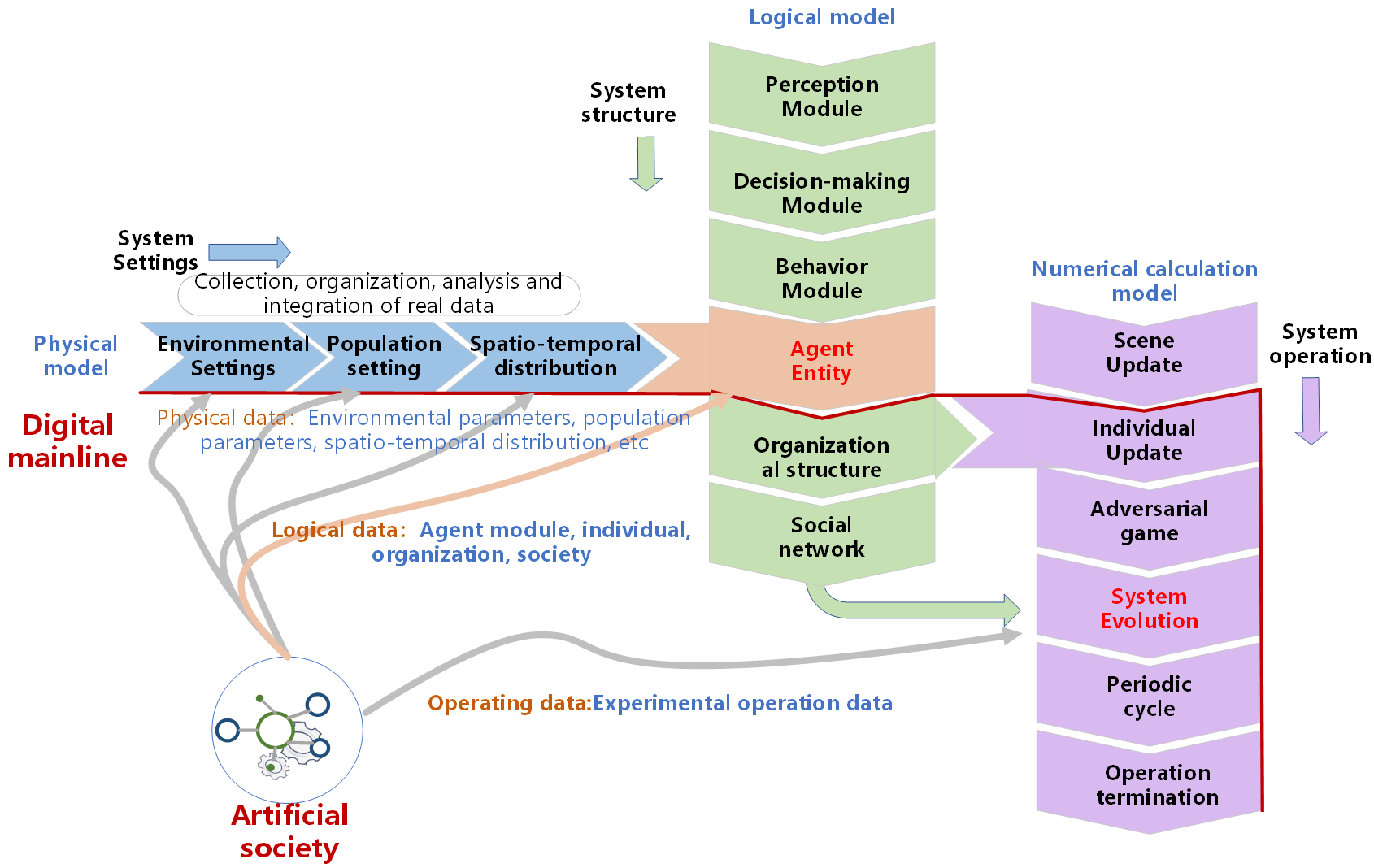}
\caption{The operation process of the artificial society.}
\end{figure*}

\subsection{Agent model}
In social simulation, an agent refers to an individual with a certain degree of autonomy, 
corresponding to biological individuals or groups in real-world societies. The agent 
individual model serves as a carrier of knowledge from various domains and can be customized 
according to the application problem, including the structure of the agent, whether it 
possesses learning capabilities, the mechanisms of interaction among agents, and so on. 
Generally, all individuals in the system adopt the same agent structure. According to 
Russell \& Norvig, agents can be classified into five types based on their perceptual 
capabilities and reasoning abilities\cite{Russell1995}:

\begin{figure*}[htbp]
    \centering
    
    \begin{subfigure}{0.35\textwidth}
        \centering
        \includegraphics[width=\linewidth]{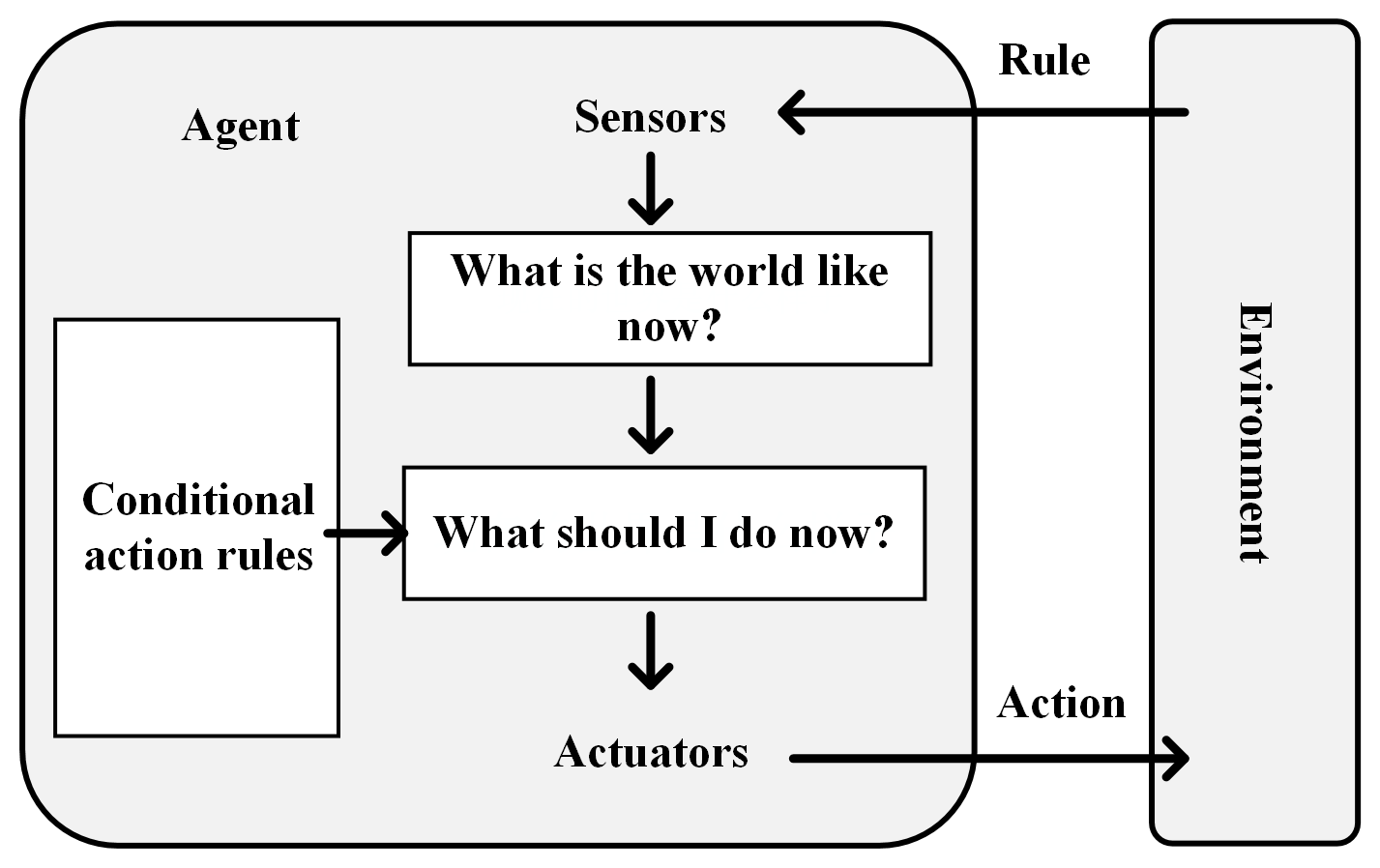}
        \caption{Simple reflex agents}
    \end{subfigure}
    \hspace{2em}
    \begin{subfigure}{0.4\textwidth}
        \centering
        \includegraphics[width=\linewidth]{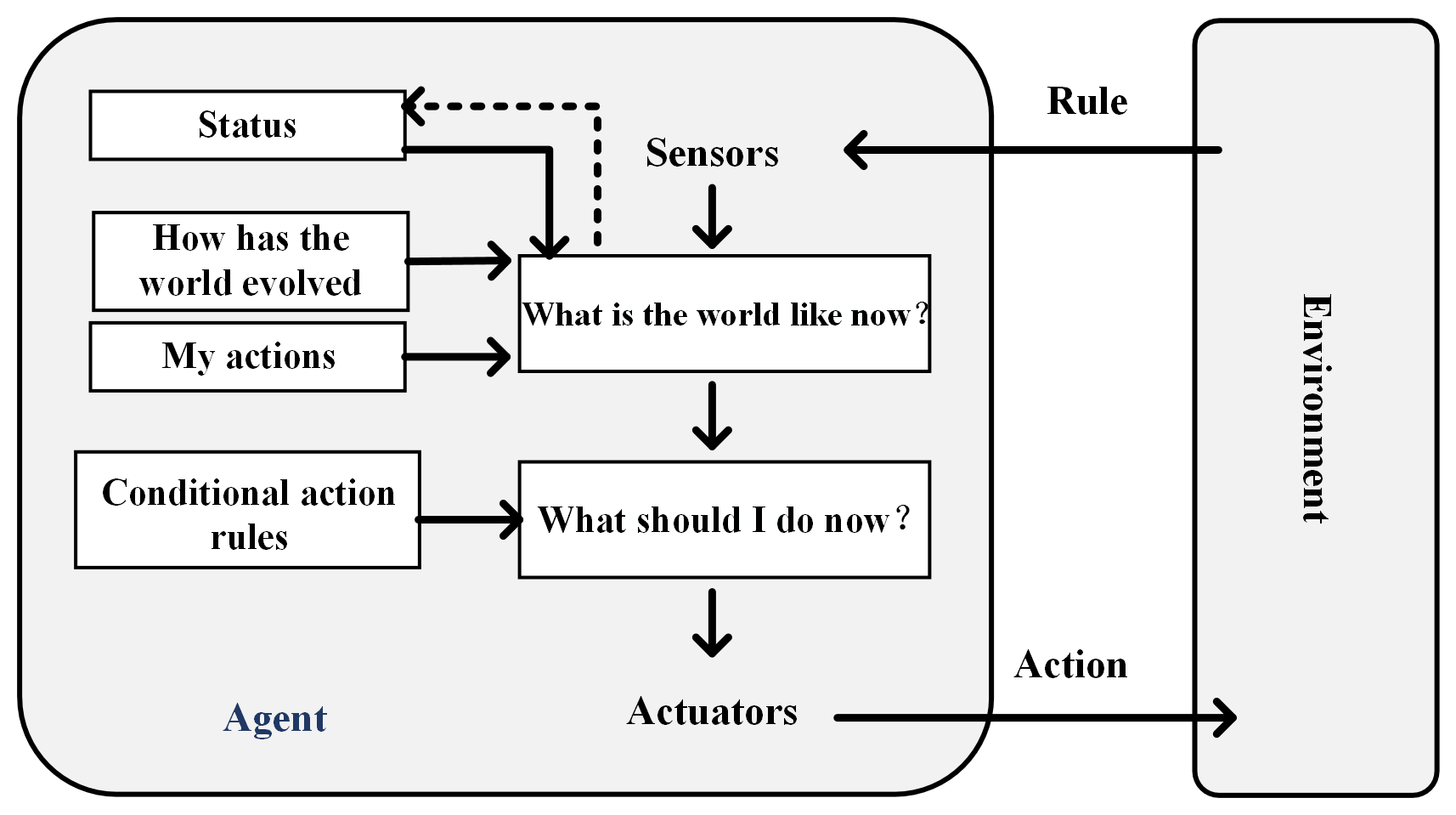}
        \caption{Model-based reflex agent}
    \end{subfigure}
    
    \vspace{1em} 
    
    \begin{subfigure}{0.3\textwidth}
        \centering
        \includegraphics[width=\linewidth]{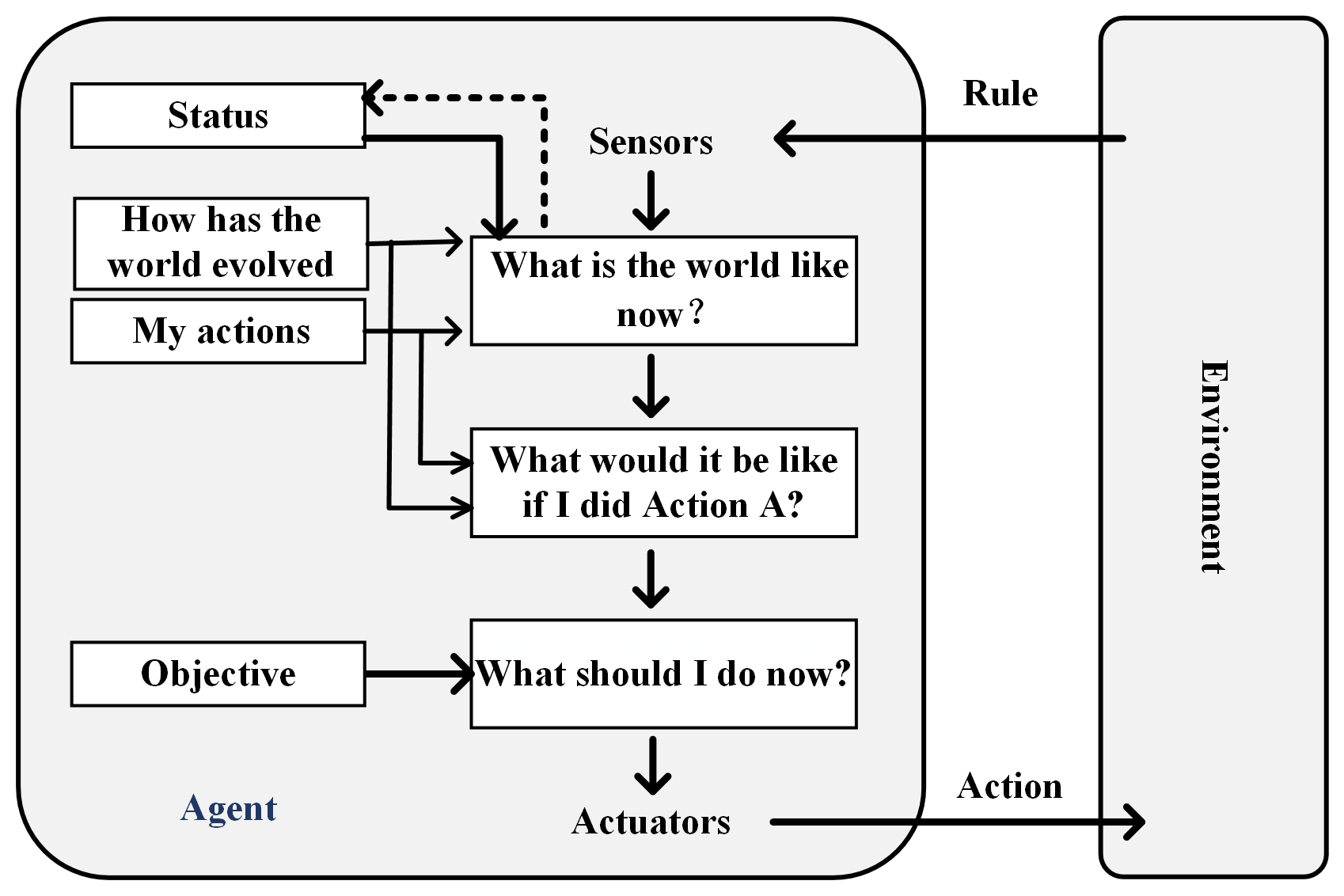}
        \caption{Goal-based agents}
    \end{subfigure}
    \hfill
    \begin{subfigure}{0.3\textwidth}
        \centering
        \includegraphics[width=\linewidth]{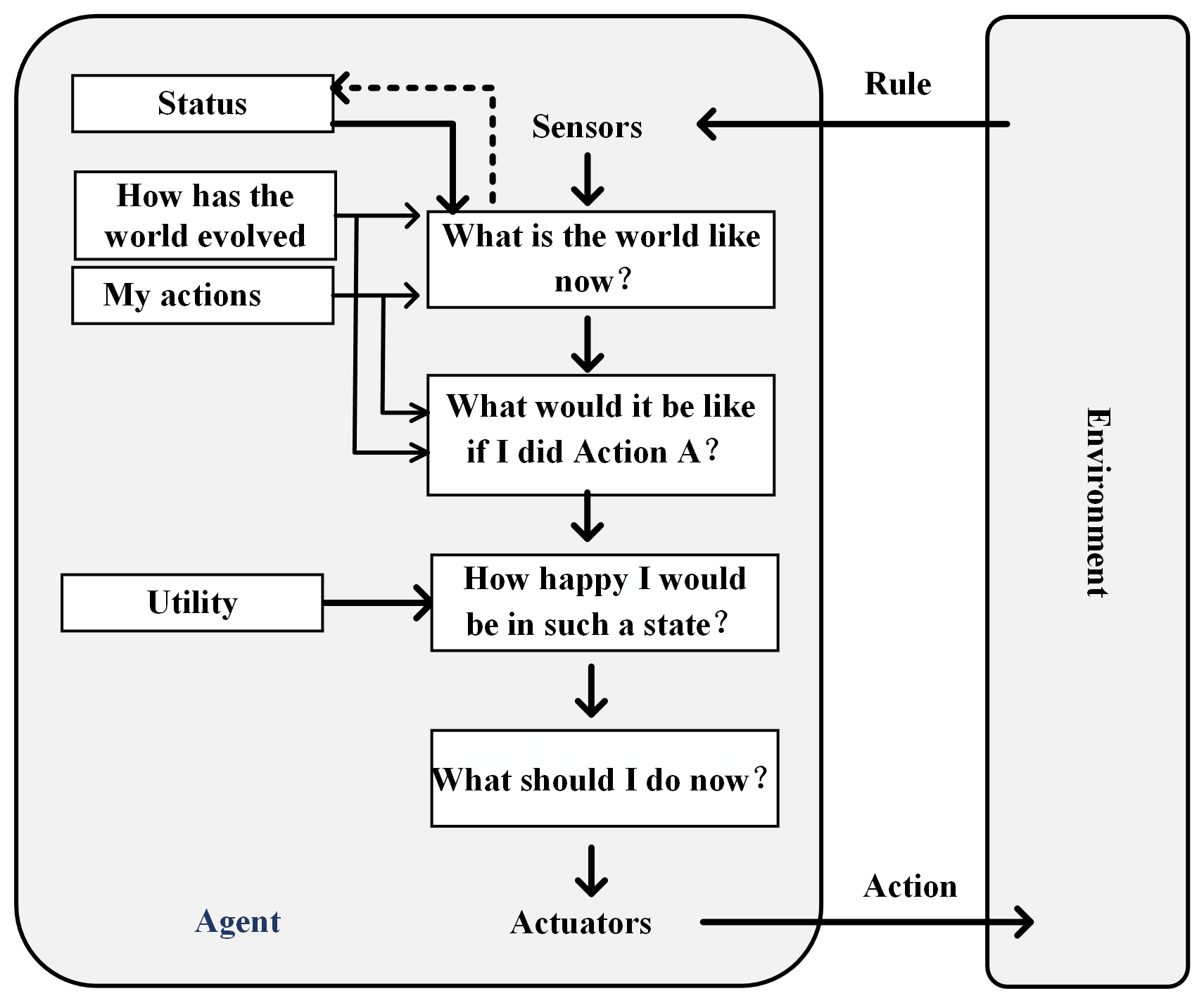}
        \caption{Utility-based agents}
    \end{subfigure}
    \hfill
    \begin{subfigure}{0.3\textwidth}
        \centering
        \includegraphics[width=\linewidth]{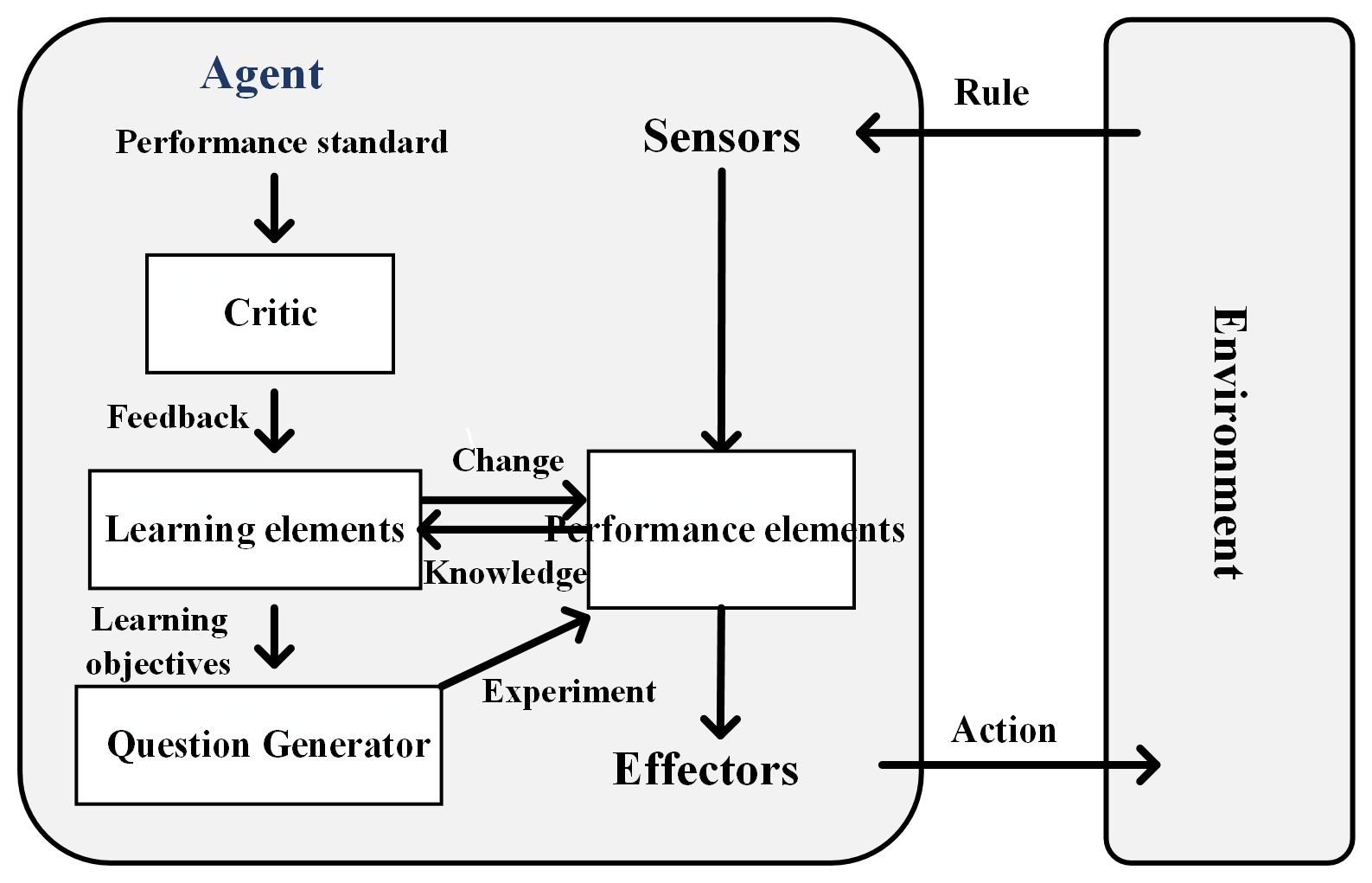}
        \caption{Learning agents}
    \end{subfigure}
  \caption{Mainstream Agent Structures}
    \label{fig:five-images}
\end{figure*}

\paragraph{Simple Reflex Agents}
Agent functionality is based on condition-action rules: “if a condition is met, then perform 
an action.” This type of agent functionality is only successful when the environment is fully 
observable. For simple reflex agents operating in partially observable environments, infinite 
loops are often unavoidable. However, if the agent can randomize its actions, it may be 
possible to escape from such infinite loops. \textbf{Figure 8(a)} shows the illustration of this type.

\paragraph{Model-based reflex agent}
An agent can operate in partially observable environments. Its current state is stored 
internally in a certain structure that describes the parts of the “unseen world.” This 
knowledge about “how the world works” is referred to as a world model. The agent can use this 
internal model to determine how its perception history and actions have affected the 
environment. It then selects the next action in the same manner as a simple agent. \textbf{Figure 8(b)} shows the illustration of this type.

\paragraph{Goal-based agents}
An agent further extends its capabilities by using “goal” information. Goal information 
describes the desired state and provides the agent with a way to make choices among multiple 
possibilities—that is, to select the path most likely to reach the goal state. Search and 
planning are subfields of artificial intelligence that focus on finding action sequences that 
achieve the agent’s goals. \textbf{Figure 8(c)} shows the illustration of this type.

\paragraph{Utility-based agents}
An agent distinguishes between goal states and non-goal states. The utility function is used 
to measure the desirability of mapping the current state to a specific state. A more general 
performance measure should allow comparison of how well different environmental states satisfy 
the agent's goals. The term utility can be used to describe the agent’s level of “happiness.” 
A rational, utility-based agent will choose actions that maximize its expected utility. The 
agent must model and track its environment, which involves extensive research on perception, 
representation, reasoning, and learning. \textbf{Figure 8(d)} shows the illustration of this type.

\paragraph{Learning agents}
The advantage of learning lies in enabling the agent to operate in unknown environments and 
acquire capabilities beyond its initial knowledge. The most important features of an Agent 
are the learning element, which is responsible for improvement, and the performance element, 
which is responsible for selecting external actions. The learning element uses a critic to 
provide feedback on the agent’s performance and determines how to modify the performance 
element or actor to perform better in the future. The problem generator is responsible for 
suggesting the next actions that will lead to new and informative experiences. \textbf{Figure 8(e)} shows the illustration of this type.

As shown in \textbf{Figure 9}, the mainstream structure of an agent currently consists of four 
components: perception, decision, reaction, and optimization\cite{Tu1994}. The information control flow 
within the agent structure connects these components into a unified whole. The \textbf{Formal 1}
expression of the Agent structure is given below, described by a set of attributes related to 
time $t$.

\begin{equation}
\textit{agent} = \langle R, S_t, E_t, Y_t, V_t, N \rangle
\label{eq1}
\end{equation}

$R$ represents the time-invariant characteristics of the agent, such as its identifier; 
$S_t$ represents the time-varying characteristics of the agent, such as its role. 
$E_t$ is the set of external events perceived by the agent that stimulate changes in its state 
and behavior. $Y_t$ is the decision-making mechanism adopted by the agent in response to 
external event stimuli and during interactions with other agents. $V_t$ is the set of 
behaviors of the agent, including both spontaneous actions and those triggered by external 
stimuli. Generally speaking, an agent follows the “benefit-increasing” principle of feedback 
learning: by continuously interacting with the environment, it gains experience in 
problem-solving and gradually optimizes its decision-making mechanism, thereby enabling it to 
take actions that are increasingly aligned with achieving its goals.

\begin{figure*}[htbp]
\centering
\includegraphics[width=15cm]{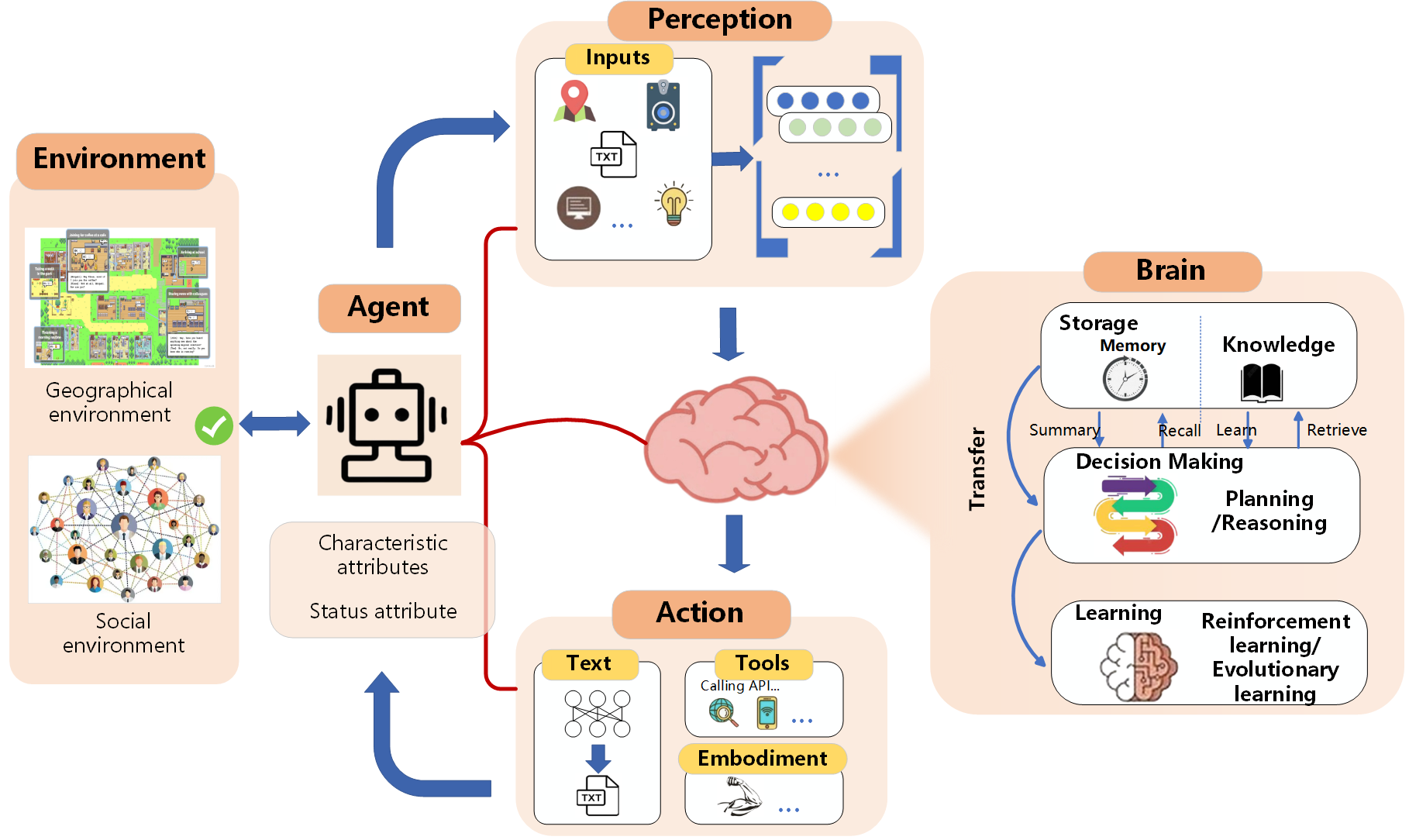}
\caption{Structural model of individual Agent.}
\end{figure*}

\subsection{Environmental model}
In an artificial society, the environmental model is the mapping of the actual physical 
environment into the computer and serves as the activity space upon which agents depend. 
According to the modeling approach, environmental models can be divided into entity-based 
modeling and grid-based modeling. Entity-based modeling refers to abstracting various 
environmental elements in the real world—such as buildings, road traffic, and climate 
conditions—into entity models. Many typical artificial society systems, such as EpiSimS\cite{Mniszewski2013}, 
are implemented in this way. Grid-based modeling does not focus on specific environmental 
objects, but rather emphasizes modeling the spatial aspects of the environment. It uses 
discrete grids to describe spatial presence and environmental attributes, such as in the 
Sugarscape model\cite{Epstein1996}.

Due to differences in the scale of artificial society scenarios, the granularity of 
environmental scene elements also varies. In large-scale artificial society scenarios—such 
as global spatial or city-level transmission analysis of diseases—the granularity of the 
environmental model is relatively coarse and generally focuses on transportation networks, 
such as airline networks. In small-scale artificial societies, the environmental model is 
relatively fine-grained and can include detailed models of natural environments, buildings, 
and road traffic. For large-scale artificial society scenarios, geographic information systems 
are typically used to build geospatial models. For small-scale artificial societies, 2D or 
3D display technologies can be used to build visual scenes, while grid technologies establish 
the geospatial coordinate system to determine the relative positions in geographic space 
(see \textbf{Figure 10}). The interaction between agents and environmental models requires attention 
to the following three aspects:

\begin{figure}[htbp]
\centering
\includegraphics[width=\linewidth]{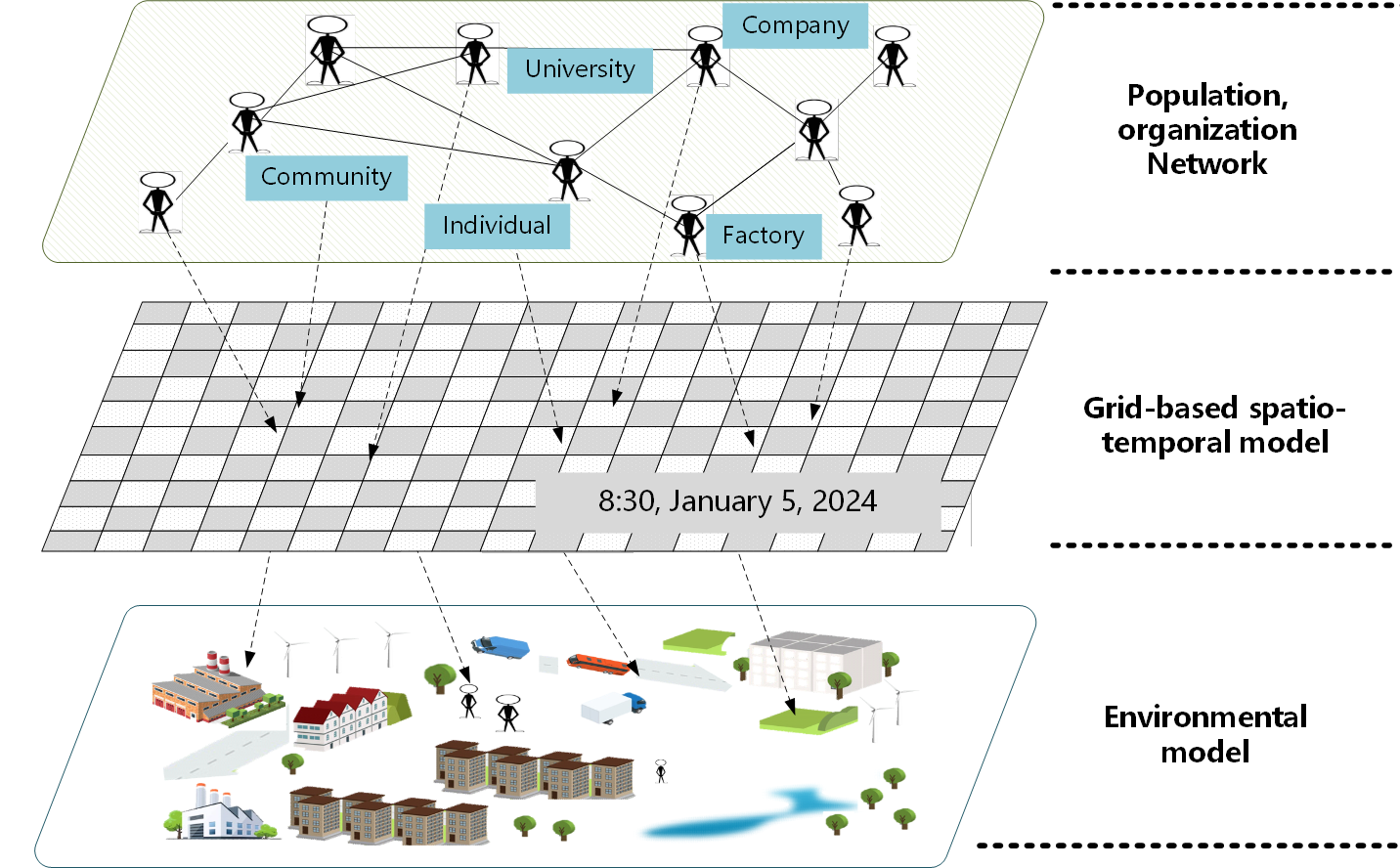}
\caption{Abstract Hierarchy of the environmental model.}
\end{figure}

\paragraph{Distribution Characteristics}
Due to practical constraints, the initial setup of environmental models is typically based on 
statistical feature data. Therefore, it is necessary to study algorithms for generating 
initialization data in artificial societies, including agents’ statistical features (total 
number, gender ratio, age distribution, etc.), agents’ geographical distribution, social 
relationship attributes of the population, statistical features of environmental entities 
(total number, type, population capacity, etc.), and the geographical distribution of 
environmental entities. The basic idea of environmental modeling is to reconstruct the 
specific features of each individual within the group from the statistical characteristics 
of group data. In this process, two aspects of consistency must be ensured: (1) ensuring that 
the number of generated agents is statistically consistent with the real world, i.e., the 
statistical characteristics of the generated agents match those of the real world; (2) ensuring 
that the internal logical structure of the generated agents is consistent with the real world, 
i.e., the organizational structure and relationships within the agents are in line with reality.

\paragraph{Spatial Correlation}
For individuals, although spatial mobility contains randomness, it more often follows 
patterns. Therefore, for each individual in the artificial society, it is necessary to 
establish association relationships with fixed activity locations. To establish matching 
associations between individuals and activity locations in a manner that conforms to human 
behavioral rules in the real world, two key tasks must be clarified: (1) determine possible 
related activity location types based on individual activity types, and establish a 
one-to-many mapping between activities and locations to ensure the agent's ability to 
randomly select activity venues; (2) search for locations that comply with travel patterns 
for each individual and match them accordingly, while optimizing the search algorithm to 
ensure that the computational time for matching large-scale populations and locations remains 
within an acceptable range. This endows individuals with behavior characteristics that reflect 
spatiotemporal correlations.

\paragraph{Spatiotemporal Correlation}
In artificial societies, agent behavior involves continuous spatial migration over time. To 
dynamically link agents with each other and with the environment, it is necessary to clearly 
define what type of activity an individual performs at what time, in what location, and what 
kind of interactions occur at that location. This ensures the temporal and spatial consistency 
and regularity of individual behavior. Three key issues need to be addressed here: (1) 
establish the mapping relationship between time and actions, i.e., define the behavioral 
temporal logic in the agent evolution process; (2)  based on behavioral temporal logic, 
integrate the regularity and randomness of agent behavior, i.e., establish a stochastic model 
of generally performed and potentially performed actions during simulation evolution; (3) 
under spatiotemporal constraints, establish a mechanism for individuals to select interaction 
partners when performing actions, i.e., clarify how to associate the social relationship 
network among individuals with behavioral models.

The development trend of environmental models is 3D digital twin visualization, which is 
closer to real-world cognition and more convenient for decision-makers to understand. In 
1994\cite{Wang2022}, Fei-Yue Wang proposed the concept of "Shadow Systems," which is very similar to the 
idea of digital twins. In 2002, Michael Grieves of the University of Michigan proposed the 
“Twinning of Systems” concept, aiming to construct digital twins equivalent to physical 
entities in virtual space, emphasizing the establishment of a bi-directional dynamic feedback 
mechanism to simulate, analyze, and optimize physical entities\cite{Grieves2005}. In 2004, Fei-Yue Wang 
proposed the concept of "Parallel Systems" as a methodology for solving complex system 
problems\cite{Wang2004ParallelSM}. In 2006, Grieves published “Product Lifecycle Management: Driving the Next 
Generation of Lean Thinking,” officially proposing the “Mirrored Space Model”\cite{Grieves2005}. In 2010, 
NASA introduced the concept of the digital twin in its Space Technology Roadmap (Area 11: 
Simulation-Based Systems Engineering): “A digital twin is an integrated, multi-physics, 
multi-scale simulation of a vehicle or system that uses the best available physical models, 
sensor updates, and operational history to mirror the life of its flying twin.”

Didi Chuxing Company is a digital space application in smart travel, which helps illustrate 
the value of digital twins. As a basic digital twin infrastructure, it mirrors the real-time 
location of all vehicles, ride requests of passengers, and all drivers' language content, 
enabling unified, scenario-based application design. Before Didi, people hailed taxis from 
the street, resulting in random matching between supply and demand. Customers had to endure 
long waits, while drivers often worried about empty mileage. The emergence of Didi broke the 
constraints of line-of-sight range, greatly reducing matching time between supply and demand, 
and allowed users to easily rate services. It can be said that digital twins have significantly 
improved both the operational efficiency and quality of the real world. \textbf{Table 4} illustrates 
the differences among digital twin companies in terms of product form, technical system, and 
value understanding.

\begin{table*}[!t]
\centering
\caption{ABM works with causal inference.}
\label{tab:4}
\begin{tabular}{|p{1.5cm}|p{7.5cm}|p{7.5cm}|}
\hline
\textbf{Type} & \textbf{Product Form} & \textbf{Understanding of Digital Twin} \\
\hline
Internet Digital Twin & Primarily PaaS platform + SDK. The PaaS platform mainly provides various case templates, reusable model resources, data panel components, and encapsulated animation APIs, data APIs, etc. Well-known domestic companies include Beijing Younuo Technology, 51World, Topology Software, etc. & Large screens, visualization, and low-code development, with a strong focus on geometric modeling, animation display, and data visualization of scenarios—that is, the twin of spatial form. The tech stack mainly revolves around WebGL or game engine technologies such as Three.js, Babylon.js, Unreal Engine, or Unity. Main application scenarios are typically in smart cities, with less involvement in industrial and equipment manufacturing sectors. \\
\hline
Industrial Software Digital Twin & Deeply rooted in the fields of automatic control and industrial simulation. Virtual scenes are built using simulation models, and real-time data is integrated to drive virtual simulation technology with real data to provide optimal processes and operational strategies for physical environments. Well-known companies include Siemens, Dassault, Schneider, etc. & Virtual products are constructed to validate product design, virtual production processes are created to validate manufacturing workflows, and parallel twins are used to verify and improve product performance. The tech stack mainly focuses on industrial simulation and IoT technologies, predominantly based on proprietary platforms. The main application areas are in industry and equipment manufacturing. \\
\hline
\end{tabular}
\end{table*}

\subsection{Rule model}
Rule models describe the cyclical mechanisms of artificial societies, including the principles 
that govern how Agents interact with each other, how environments interact, and how agents 
interact with the environment. These rules can either be mappings of real-world social rules 
or artificially assumed hypothetical rules. As shown in \textbf{Figure 11}, artificial societies 
adopt a feedback loop approach to establish a three-level evolutionary order:

\paragraph{Individual Evolution}
At the bottom level is the individual evolution space, which simulates the phenomenon of 
individuals undergoing learning and evolution within a social system. In a social system, 
different individuals have different interests and needs, and individual interests may not 
align with the overall interests of the system, or may even conflict. As a result, behaviors 
such as “countermeasures to top-down policies” often arise, which affect the implementation 
and effectiveness of policies. Cooperation among agents is not always smooth; conflicts may 
occur, including resource conflicts arising from multiple Agents competing for the same 
limited resource; conflicts where one party’s goal realization prevents the other from 
achieving theirs; and conflicts where different Agents reach contradictory conclusions on 
the same issue.

\paragraph{Group Emergence}
If individual behavior is regarded as an input causal variable, then emergence appears as 
the output phenomenon. When individual behaviors change in response to system feedback, the 
system may evolve from first-order emergence to second-order emergence. In this way, a 
closed-loop is formed between the overall system and individual behavior—namely, a 
bidirectional feedback mechanism (feedforward and feedback), which may lead to positive 
feedback loops and mutual reinforcement. Under such conditions, even a small deviation from 
equilibrium can lead to large and exaggerated effects.

\paragraph{Feedback Regulation}
The regulation mechanism refers to a mechanism by which macro-level phenomena exert 
adjustments on micro-level individual behavior. An effective regulation mechanism must 
consider the individual’s pursuit of interests, to ensure the system’s motivation and 
vitality for development; at the same time, it must firmly prevent interest imbalance to 
avoid threatening the system’s order and stability. To ensure that groups composed of multiple 
agents can act in a coordinated manner, coordination becomes the core issue in multi-agent 
systems. Coordination is the management of interdependencies among activities, including the 
appropriate arrangement of their goals, resources, and mental states. It is reflected in 
shared resource constraints, simultaneity constraints, producer/consumer relationships, and 
task/subtask relationships.

\begin{figure}[htbp]
\centering
\includegraphics[width=\linewidth]{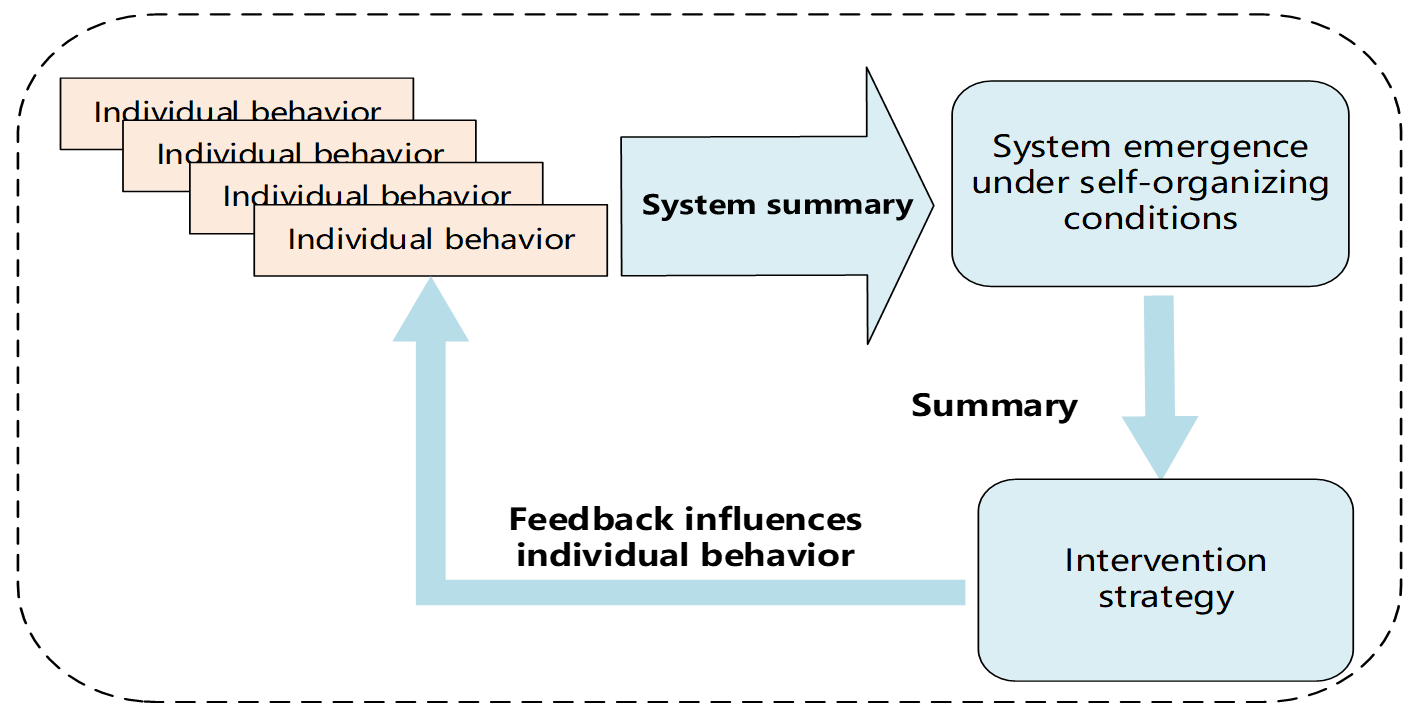}
\caption{The operational logic framework of the rule model (replaced with game regulation).}
\end{figure}

The mutual nesting and causality of the evolution mechanisms among various levels make this 
analytical framework exhibit the characteristics of complex system theory. These three levels 
and their interactions constitute a relatively complete and abstract circular analytical 
structure of the evolution process of artificial society. Each level focuses on different 
aspects of the socio-economic system (such as individuals, organizations, clusters and 
countries etc.), and the relevant modeling techniques can be selected according to specific 
needs. The implementation logic of the rule model is shown in the \textbf{Algorithm 1}.

\begin{algorithm*}[htbp]\small
\caption{Implementation logic pseudo-code of the rule model}
\label{alg:rule-model}

\begin{algorithmic}

\STATE \textbf{Definition 1 -- Knowledge Point:} where $X_i$ represents the experience possessed by the $i$-th agent, and $K_j$ represents the $j$-th part of that experience.

\STATE \textbf{Definition 2 -- Knowledge Space:} The set of all possible knowledge points.

\STATE \textbf{Definition 3 -- Agent:} An entity representing an individual within the artificial social system.

\STATE \textbf{Step 1: Individual Evolution} \\
In the real world, individuals must continuously enhance their abilities to gain a competitive edge in fierce competition. Thus, variation is the generative mechanism of diversity and the driving force behind the evolution of social systems. Without variation and innovation, evolution does not exist. In the modeling framework, the rules of variation and evolution for agent individuals can be defined using various machine learning or evolutionary algorithms. The simulation process emphasizes three characteristics: first, all variation processes are fundamentally uncertain, including both the random alteration of individual strategies and the generation of new strategies; second, variant individuals must be capable of acting quickly to reap benefits before other actors complete imitation; third, variation is subject to numerous constraints at various levels, including various risk factors it may introduce.

\STATE \textbf{Step 2: Organizational Evolution} \\
The selection mechanism is the mechanism that reduces diversity. It evaluates the adaptability of individuals based on certain criteria, selects highly adaptive evolutionary units, and eliminates poorly adaptive ones. Without a selection mechanism, the social system would lose the ability to distinguish between good and bad, leading to a divergent, chaotic, and inefficient state. Therefore, a sound selection mechanism is a critical filtering process in the evolution of artificial society. Since defining fitness as a criterion for individuals to produce offspring (or undergo genetic replication) is essentially the same as the selection process, the evolution rules for organizations in the modeling framework can be defined according to evolutionary features of biological communities, such as Ant Colony Optimization (ACO)\cite{Dorigo2007}, Particle Swarm Optimization (PSO)\cite{Kennedy1995}, Artificial Bee Colony (ABC) algorithm, and others\cite{Karaboga2009}.

\STATE \textbf{Step 3: Social Emergence} \\
After intense competition, some elites stand out from the individual pool. Other individuals can enhance their survival capabilities in the ecosystem by imitating and learning from these elites—this marks the phase of social evolution. The social evolution process involves the diffusion of elite characteristics and the formation of culture. There are three typical diffusion models: the contagion model\cite{Dodds2005}, the social threshold model\cite{Chen2010}, and the social learning model\cite{Hoppitt2013}. Factors influencing diffusion mechanisms also include: (1) the structure of the interaction network. In regular network structures, interaction clustering among individuals is high but diffusion speed is low. In random network structures, interaction clustering is low but diffusion speed is high. (2) the degree of heterogeneity among interacting agents (including differences in preferences, cognition, beliefs, knowledge stock, learning ability, and absorptive capacity). Generally, the greater the heterogeneity, the slower the diffusion speed. (3) organizational rules. Different organizational rules shape different interaction patterns, thus influencing the speed of diffusion of variation.

\STATE \textbf{Step 4: Feedback Regulation} \\
The social space, in turn, acts upon the individual space—this is the mechanism by which macro-level phenomena affect micro-level individuals. In the real world, cooperation and competition among individuals shape the cultural atmosphere of the entire social ecosystem, which in turn influences individuals’ decision-making mechanisms. To simulate the phenomenon that culture can accelerate individual evolution, feedback rules are designed in the modeling framework. These influences may be reflected in the following aspects: (1) an effective selection mechanism enables highly efficient individuals to acquire more resources, facilitating further technological accumulation and innovation; (2) the selection mechanism itself shapes the cognition, preferences, and internal constraints of variant individuals, thereby exerting systemic influence on variation; (3) the selection mechanism of intervention strategies not only affects the variation level of specific individuals, but also that of the entire system. Replicator dynamics is a typical deterministic and nonlinear evolutionary game model based on the selection mechanism. Adding random strategic variation to this model results in a composite evolutionary game model that includes both selection and variation mechanisms, commonly referred to as the replicator–mutator model.

\STATE \textbf{Step 5: The Next Cycle} \\
As time progresses, some elites may fall behind, while newly capable individuals may rise to become the new elites. Ultimately, the evolutionary equilibrium of the entire social system is broken, and a new cycle begins.

\end{algorithmic}
\end{algorithm*}

\section{The classic cases of ABM}
The application scope of social simulation includes three aspects: phenomenon interpretation, trend prediction, and strategy optimization. This section presents some classic cases in these three aspects to help readers deepen their understanding of social simulation.

\subsection{Thought Experiments}
Thought experiments do not model specific scenarios or real-world social systems, but rather describe the abstract logical relationships of general social systems. They aim to explore and quantitatively analyze the unpredictable consequences that certain assumptions may bring to human society through experiments. This type of research avoids the mapping issue from the actual society to the artificial society, typically offering metaphors, inspirations and qualitative trends rather than precise solutions to complex issues.

\paragraph{Sugarscape model}
In 1996, Joshua Epstein and Robert Axtell proposed an "artificial society" model, Sugarscape, which can conduct relevant experiments in economics and other social sciences\cite{Epstein1996}. As shown in \textbf{Figure 12}, the Sugarscape model is a closed world composed of squares: the red dots represent the Agent, which can only move around in this world; the yellow part represents social wealth - sugar, and the intensity of yellow indicates the concentration of sugar distribution. Each agent contains three attributes: visual range r, the resource metabolic rate v, and the amount of sugar s it possesses. Agents move around according to the following rules: (1) Observe all cells within the visual range r and determine the cell with the highest sugar content; (2) If there is more than one cell with the maximum sugar content, choose the nearest one. (3) Move to this square; (4) Collect the sugar in this cell and update the corresponding variable.

\begin{figure}[htbp]
\centering
\begin{minipage}[t]{0.2\textwidth}
  \centering
\includegraphics[width=\linewidth]{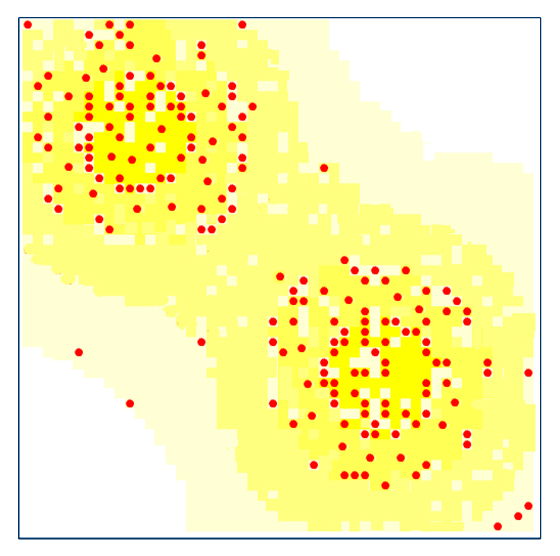}
\caption{Distribution of sugar and Agent in Sugarscape.}
\end{minipage}
\begin{minipage}[t]{0.2\textwidth}
  \centering
\includegraphics[width=\linewidth]{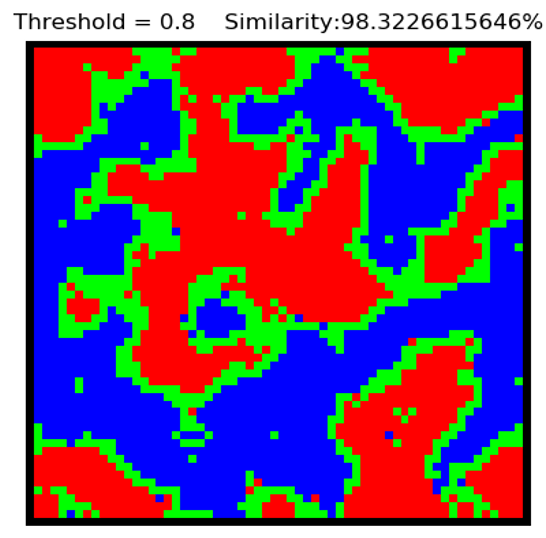}
\caption{Racial segregation Experiment Based on the Schelling model.}
\end{minipage}
\end{figure}

As shown in \textbf{Figure 11}, as the experiment progresses, some agents—due to their superior individual abilities or advantages in local resource distribution—are able to collect more sugar; meanwhile, individuals lacking sugar gradually wither and die, resulting in most agents clustering in the two regions with higher sugar concentrations. Eventually, a small number of Agents possess a large amount of sugar resources, while the majority own only a small amount, thereby confirming the well-known Matthew effect in social science. By adding multiple types of resources (e.g., spices) to the Sugarscape artificial society, researchers can study how individuals in real societies form markets through resource exchange. By altering the behavioral rules followed by agents, this model can be used to explore social phenomena such as environmental change, genetic inheritance, trading interactions, and market mechanisms. The Sugarscape model provides insights into wealth inequality, market formation, and other social phenomena by simulating agent interactions around resource collection.

\paragraph{Schelling Segregation model}
The Schelling Segregation Model, also known as the Schelling segregation model, was proposed by American economist Thomas Schelling\cite{Schelling1971}. This model illustrates the impact and role of homogeneity in spatial segregation, revealing the underlying mechanisms behind racial and income segregation. The model consists of three key elements: agents who generate behavior, the behavioral rules these agents follow, and the macro-level outcomes resulting from agent behavior. As shown in \textbf{Figure 13}, the experiment conceptualizes an entire city as a giant checkerboard, where each grid cell can be either occupied by an agent or remain vacant. There are two types of agents (red and blue), with approximately equal numbers, and around 10\% of the cells are left empty (green). Each agent has a minimum threshold—if the proportion of similar neighbors falls below this threshold, the agent will relocate to an unoccupied cell that meets its residential preference. The behavioral rules of the agents are as follows: (1) Count the number of same-type neighbors nearby; (2) If the number of similar neighbors is greater than or equal to the preference threshold, the Agent is satisfied and stops moving; otherwise, it continues moving; (3) The agent will search for the nearest unoccupied cell that satisfies its threshold and move to that location.

Over time, the degree of segregation between different types of Agents will become increasingly pronounced. By modifying the experimental settings—such as agent lifespan, the amount of empty space, and so on—it becomes evident that changing tolerance alone is not sufficient to prevent segregation, as nearly every agent prefers all neighbors to be of the same type. This phenomenon raises deeper questions about social issues: To what extent must racism persist before the entire societal structure shifts toward such a segregated pattern? How can racial segregation be reversed? Researchers have extended this type of model and applied it to studies of ethnic conflict in different regions, yielding encouraging results\cite{Markisic2012, Bhavnani2008}.

\paragraph{Landscape theory}
Robert Axelrod of the University of Michigan proposed the landscape theory to study alliance formation in human societies and to predict the equilibrium states they are likely to reach\cite{Axelrod1993}. This theory treats social units as entities that are drawn together by attractive forces and pushed apart by repulsive forces. This push-and-pull environment causes particles to cluster into alliances, and the final alliance configuration arranges these entities in the most stable manner. To describe this more effectively, Axelrod and his colleagues defined the concept of “total energy”: for each possible configuration, the total of all attractive and repulsive forces between group members is calculated. The energy landscape is a graphical representation of all possible combinations and their corresponding energy levels—the more stable a combination, the lower its energy. Axelrod and his team used the landscape model to simulate potential alliance blocs on the eve of World War II, using six factors as the basis for determining interactions among countries: ethnic composition, religious belief, territorial disputes, ideology, economic conditions, and historical ties. Each factor was assigned a simple weight: +1 for convergent (cooperative) factors and –1 for conflictual or hostile factors.

\begin{figure}[htbp]
\centering
\includegraphics[width=\linewidth]{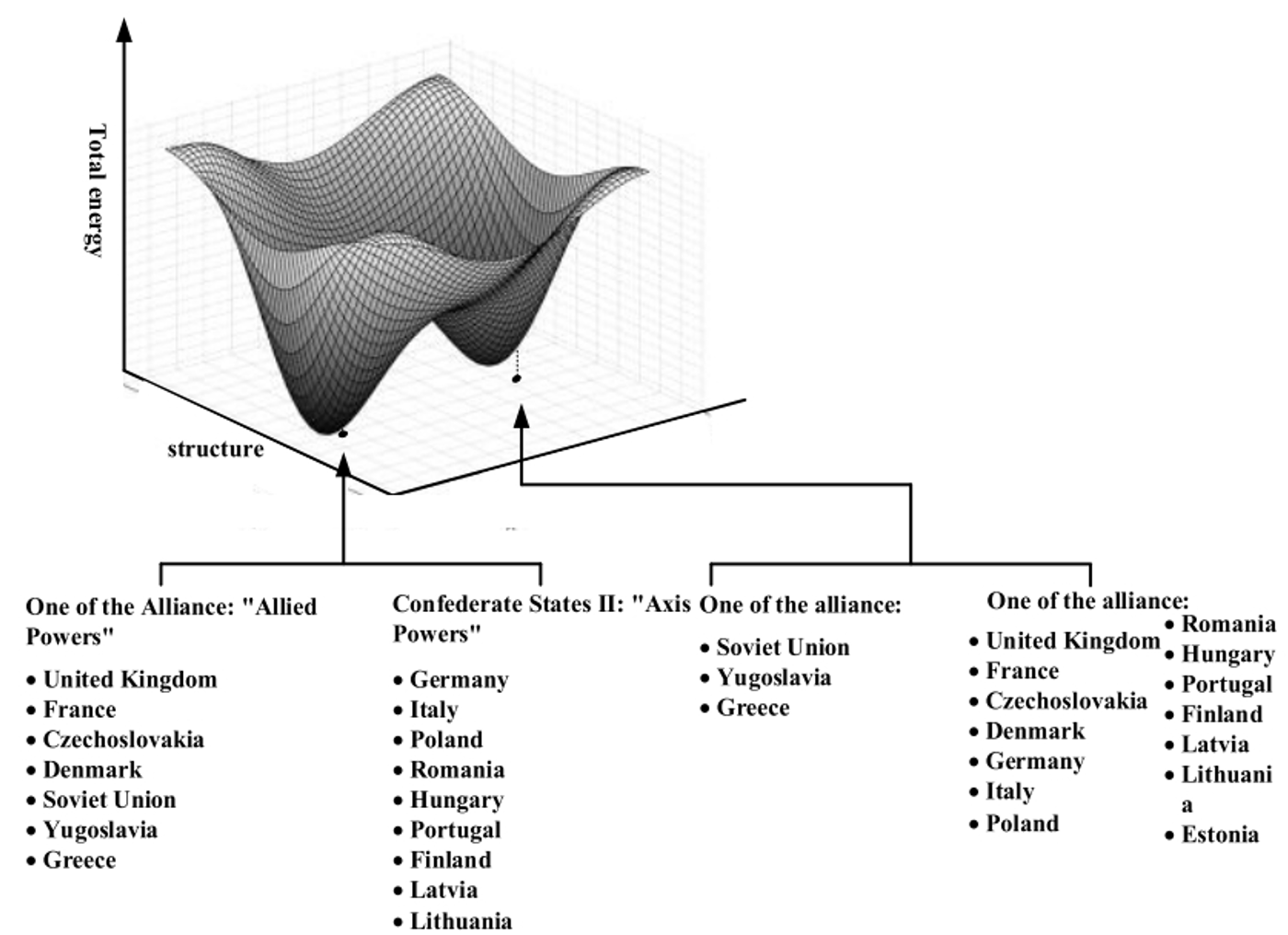}
\caption{A picture of the Alliance camp on the eve of World War II.}
\end{figure}

\textbf{Figure 14} sketches the scenarios of all the combinations. There are two basin structures in the figure, namely the energy minima. The one that sinks deeper is very close to the actually formed alliance of the Central Powers and Axis Powers. Only Portugal and Poland have been classified into the "wrong" camp. Another basin foretold a very different scenario: the Soviet Union was in confrontation with all the other European countries. According to this energy landscape map, which "trough" Europe will eventually fall into is determined by where the starting point is located. The landscape theory tells us that for the special case of an alliance, it is possible to conduct actual calculations and obtain values with a certain degree of credibility, and the results are very similar. Through the calculation and experimental deduction of the landscape model, historical research becomes more concrete. When discussing the world situation, with a certain degree of quantitative language, the influencing factors in the historical development process can be identified, clarifying their scope and thus achieving the purpose of prediction.

\subsection{Mechanism exploration}
Mechanism exploration focuses on modeling real-world social systems, emphasizing a high level of alignment between artificial social systems and real-world social systems, aiming to address existing or potentially emerging problems in society. Such applications often face the challenge of validating the accuracy of the mapping from the real society to the artificial one. Current researchers primarily aim to address this issue through rapidly evolving data acquisition and processing technologies, thereby enabling experimental results to address real-world questions.

\paragraph{Artificial stock market (ASM)}
The stock market is clearly a complex system. Its multiple components and hierarchical structure make the stock market as a whole exhibit ccharacteristics of complexity and unpredictability. To explore and understand how investors make portfolio choices, the Santa Fe Institute (SFI) proposed the "Artificial Stock Market" model (ASM) in 1987\cite{arthur2018asset}. It aims to understand the complexity of economic system from a novel perspective. This model abandons the assumption of a completely rational, omniscient, and omnipotent "economic man", and instead employs a bounded rational individual capable of learning, adapting to the environment and making decisions through induction. It regards the economic system as a complex system in which several interacting individuals exist. In this virtual market, several trading agents make predictions by observing the dynamic changes in stock prices and dividends in the digital world. Based on these predictions, they decide whether to purchase stocks and the quantity to buy, in order to maximize their own utility. All agents independently form their own expectations and have the ability to learn, allowing them to adjust their decisions based on the success or failure of their previous predictions. Conversely, the decisions made by all traders determine the state of supply and demand competition, and thereby determine the price of stocks.

\begin{figure}[htbp]
\centering
\includegraphics[width=\linewidth]{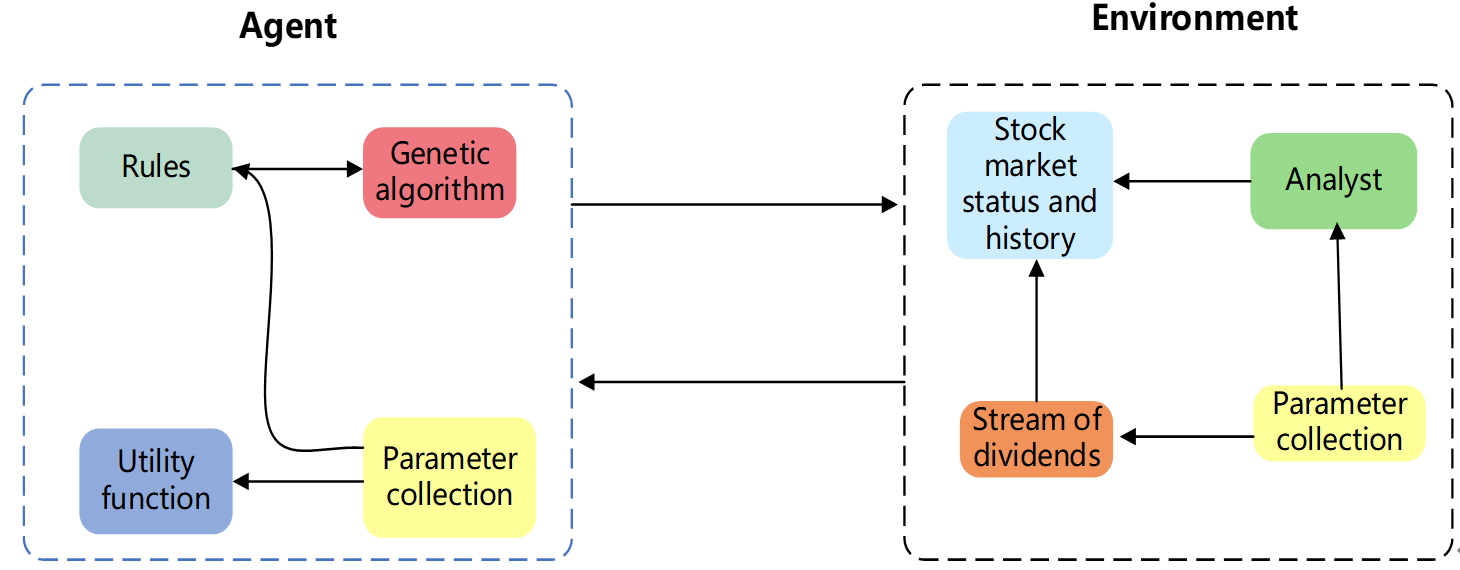}
\caption{Structure of the interaction between agent and the stock market.}
\end{figure}

As shown in \textbf{Figure 15}, the entire stock trading market constitutes a self-contained computational system. The model operates in discrete time steps, and the experiment can proceed indefinitely. At the beginning of each cycle, the current dividend is issued and perceived by all agents. Each agent uses current and historical data on dividends and prices to match with the “condition” part of their rules and forms expectations for the next period’s price and dividend. Based on these expectations, they then optimize the allocation of their funds between stocks and bonds. A market maker calculates the new stock price according to the aggregate supply and demand imbalance of all agents and a market-clearing price equation. At the start of the next time step, agents update the accuracy of their prediction rules, and the new dividend is distributed. This process repeats until a termination condition is satisfied. The ASM (Artificial Stock Market) offers a powerful metaphor for real-world stock markets. It can be used to test whether different expectations eventually evolve into identical rational expectations, thereby supporting the Efficient Market Hypothesis, or whether more complex behaviors emerge at both the individual and organizational levels, thus supporting the perspective of real investors and helping to explain real-world financial market phenomena. Subsequently, many researchers have improved the ASM model to conduct deeper analyses, such as the link between micro-level investor behavior and macro-level stock market dynamics, or the effectiveness of price limits.

\paragraph{Political mechanism}
Claudio Cioffi-Revilla from George Mason University proposed the MASON RebeLand model to analyze how the state of a polity or its political stability is influenced by internal (endogenous) or environmental (exogenous) processes, such as changes in the polity's economy, population, culture, natural environment (ecosystem), climate, or broader socio-natural stressors\cite{Cioffi-Revilla2010}. \textbf{Figure 16} presents a “map” view of RebeLand as a polity or nation situated within a natural environment. The country consists of an island surrounded by water, thus omitting interactions with external or neighboring countries. The RebeLand environment includes terrain and a simplified weather system that can simulate climatic dynamics (e.g., prolonged droughts, climate change, etc). The rebellion and political elements include a society and a government system that addresses public issues through policy-making. Initially, the government formulates policies to address social problems. In subsequent simulations, under certain conditions, the society may also give rise to rebels who interact with the government's military, along with other emergent phenomena.

\begin{figure}[htbp]
\centering
\includegraphics[width=\linewidth]{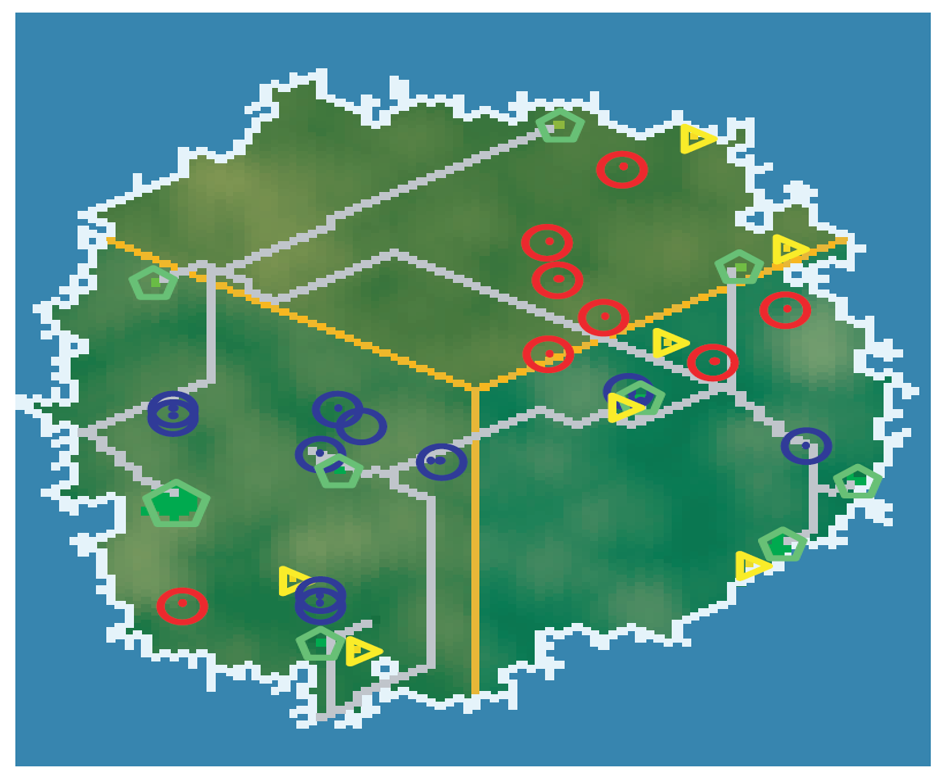}
\caption{The map of RebeLand Island features its main natural and social characteristics. Cities are indicated in green, natural resources in yellow, and the rebel forces and government forces in red and blue respectively. Roads and provincial boundaries are represented by grey and yellow respectively. The physical topography is displayed in green tones, and the island is surrounded by the ocean.}
\end{figure}

There are two types of rebels. The main agents include the general population, cities and countries. Cities represent local public administrative organizations, while the state represents the entire government system of the polity. Secondary agents include rebels, rebel groups and government forces representing police and military units that arise from among the general population under various conditions. The rebels are supported by rebel groups, which fund alternative policies that oppose the official policies. Therefore, rebel groups represent the alternative political organizations that compete with the state. The government forces attempt to destroy the rebels by attacking them. This study presents three scenarios, showing respectively stable, unstable and failed political states. An important outcome is related to the overall resilience of the political system. Usually, the failure of a political system does not result from one or a few pressure issues to experience, but rather a large set of issues (such as inflation, rebellion and environmental pressure). Such results may not immediately lead to actionable policy recommendations for specific projects, but at least they could provide valuable new insights for researchers and policy analysts. This type of research is also used in election prediction and applied political science research\cite{Laver2011}.

\paragraph{Epidemic spread}
With the outbreaks of emerging infectious diseases such as SARS, H1N1, and COVID-19, epidemic transmission models have become a research hotspot. Traditional predictive models rely heavily on variable selection and expert knowledge, attempting to make optimal estimates based on uncertain information. However, the spread of infectious diseases is a dynamic and uncertain process, involving numerous influencing factors as well as many unknown aspects and unexpected events, making predictive research extremely challenging. Take the infection rate of a virus as an example—it not only depends on local social behavior, public health conditions, and political decisions, but also changes in response to evolving intervention measures. Computational experiments use artificial society models to achieve integrated analysis of basic data, modeling methods, and analytical results, mainly involving three aspects: the generation of epidemic dynamics (e.g., initial number of infections, contact rate, transmission speed, virus incubation period, mortality rate, recovery probability, etc.); the representation of spatial and geographical features (e.g., city type, transportation networks, population density, temperature, weather, age distribution, urban infrastructure, etc.); and the modeling of resources and governance capacity (e.g., medical resources, number of hospital beds, social organizations, control measures, information transparency, etc.). Different artificial society systems vary in their implementation methods, representation techniques, and levels of accuracy. \textbf{Table 5} presents a comparative summary of the characteristics of several artificial society systems related to infectious disease research.

\begin{table*}[!t]
\centering
\caption{Comparison of Typical Characteristics of Large-scale Propagation Simulation Systems.}
\label{tab:5}
\begin{tabular}{|p{2cm}|p{2cm}|p{2cm}|p{2cm}|p{2cm}|p{2cm}|p{2cm}|}
\hline
\textbf{Feature} & \textbf{BioWar\cite{Carley2006}} & \textbf{EpiSimS\cite{Mniszewski2013}} & \textbf{GSAM\cite{Epstein2009}} & \textbf{CovidSim\cite{Schneider2020}} & \textbf{ASSOCC\cite{Ghorbani2020}} & \textbf{SIsaR\cite{Pescarmona2020}} \\
\hline
\textbf{Disease Type} & Droplet transmission, physical contact & Smallpox, influenza & Not specific to any disease, uses H1N1 as an example & COVID-19, other respiratory viruses & COVID-19 & COVID-19 \\
\hline
\textbf{Main Purpose} & Impact assessment, strategy optimization & Impact assessment, strategy optimization & Study the spread and control of infectious diseases & Impact assessment, strategy optimization & Impact assessment, policy trade-off & Assess costs and benefits of different intervention policies \\
\hline
\textbf{Simulation Scale} & Medium-sized U.S. city & Medium-sized U.S. city & Global & National & National & National \\
\hline
\textbf{Simulation Method} & Multi-agent & Multi-agent & Multi-agent & Geospatial Unit & Multi-agent & Multi-agent \\
\hline
\textbf{Programming Language} & C++ & - & Java & C++ & R language \& NetLogo & NetLogo \\
\hline
\textbf{Visualization} & NO & YES & YES & YES & YES & YES \\
\hline
\textbf{Open Source} & NO & NO & NO & YES & YES & YES (available online) \\
\hline
\end{tabular}
\end{table*}

At present, computational experiments have become an important means of studying large-scale infectious disease transmission, mainly applied in three fields: (1) Trend forecasting. The development of transportation systems has made it easier for major infectious diseases to spread on a large scale. Analyzing and predicting their transmission trends before reaching a specific area is a key prerequisite for emergency preparedness. (2) Impact pre-assessment. The outbreak and evolution of an epidemic within a region pose direct threats to public health and cause secondary effects on the socio-economic environment. Quantitative assessment of its impact serves as an important basis for emergency stockpiling and intervention decisions. (3) Intervention strategy optimization. A variety of intervention measures can be used in emergency prevention and control of major infectious diseases, each with different targets and intensities—for example, the choice between targeted immunization and herd immunity. In practice, selecting reasonable interventions and their intensities to form a combined intervention strategy—one that controls the epidemic while minimizing the cost of intervention—is a major challenge in emergency decision-making. However, no computational model exists that perfectly matches the real-world transmission process. All models are merely approximate abstractions of the actual transmission process and must continuously avoid and reduce uncertainty in data, modeling, computation, and result analysis in order to achieve simulation outcomes that increasingly approximate the real transmission process.

\subsection{Parallel optimization}
The purpose of scientific research is to provide a causal hypothesis that can link existing facts. If the constructed artificial society model represents such a regularity, then the output behavior of the model can exert an effective impact on the real world. The parallel world, by establishing an artificial society model that is homomorphic to the real society, enables parallel execution and cyclic feedback between the two, thereby supporting the management and control of real complex systems.

\paragraph{Island economy}
Economic inequality is worsening globally and has drawn increasing attention due to its negative impacts on economic opportunity, health, and social welfare. Governments can use economic policies—particularly in taxation and redistribution—to improve social outcomes. However, the challenge lies in the fact that taxation tends to reduce productivity. Workers may choose to withdraw from labor if they are required to pay taxes on their income, thereby decreasing the utility of labor. This effect can be especially significant for highly skilled workers. Thus, a trade-off must be made between economic equality and productivity: do redistributive interventions that allow wealth to be reallocated result in a reduction of total wealth available for redistribution? Because of the coupling relationship between taxation and labor, the key unresolved issue is how to mitigate economic inequality while maintaining productivity. In practice, such economic problems are difficult to study due to the lack of suitable economic data and the limited opportunities for experimentation.

\begin{figure}[htbp]
\centering
\includegraphics[width=\linewidth]{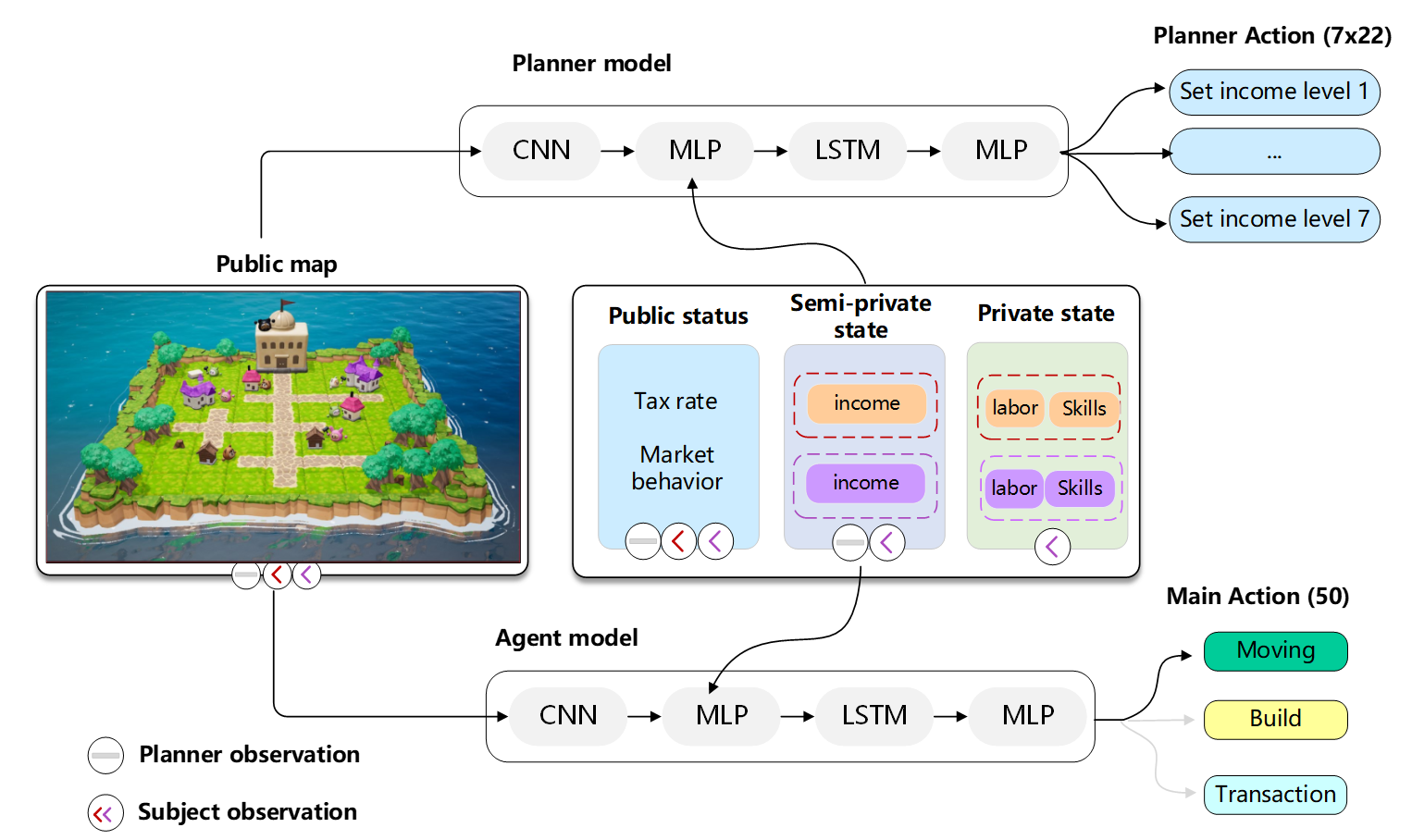}
\caption{Island Economy.}
\end{figure}

To this end, Salesforce proposed a new research project called "The AI Economist". It conducts economic simulation through AI agents, where both individuals (agents) and tax strategies (Planners) possess learning and adaptive capabilities\cite{zhang2021synergistic}. This project introduces a two-level reinforcement learning framework involving citizens and tax policies, aiming to discover tax strategies that efficiently balance economic equality and productivity. This framework operates without relying on prior knowledge or modeling assumptions, enabling direct optimization for any socioeconomic objective and learning solely from observable data. \textbf{Figure 17} presents the core implementation framework of The AI Economist, focusing on the evolutionary game mechanism between agents and the Planner—that is, how reinforcement learning is used to jointly optimize individual agent behavior and the tax policy employed in the economy.

In the inner loop, various types of worker agents gain experience by performing labor, earning income, and paying taxes. They adapt their behavior through trial-and-error learning to maximize their individual utility. In the outer loop, the Planner agent adjusts the tax policy to optimize a social objective, which is based on a social welfare function from economic theory that considers the trade-off between income equality and productivity. The AI Economist offers a powerful framework that integrates economic theory with data-driven simulation, producing objective insights into the economic consequences of different tax policies. However, there are still some limitations—for example, it does not yet simulate human behavioral factors or interpersonal interactions, and it focuses on a relatively small economy. The developers at Salesforce hope that The AI Economist can address the complexities that traditional economic research struggles to handle and provide an objective tool for studying the real-world impact of policies. Researchers can further refine this framework for use in “what-if” experiments and counterfactual reasoning, which is often a key value-added feature of AI-driven simulators.

\paragraph{Virtual Taobao}
Product search is the core business of Taobao, which is one of the largest retail platforms. Taobao’s commercial objective is to increase sales revenue by optimizing the strategy of displaying PVs (page views) to customers. Since the feedback signal from customers depends on a sequence of PVs, the problem can be viewed as a multi-step decision-making problem rather than a single-step supervised learning task. The engine and the user serve as each other’s environment. RL (Reinforcement Learning) solutions excel at learning sequential decision-making and maximizing long-term rewards. However, one major obstacle in directly applying RL in these scenarios is that current RL algorithms typically require a large number of interactions with the environment, which incurs high physical costs—such as spending real money, taking days to months, poor user experiences, or even life-threatening risks in medical tasks.

To avoid physical costs, researchers use a simulator (i.e., Virtual Taobao) to conduct reinforcement learning training, allowing policies to be trained offline in the simulator using any RL algorithm that maximizes long-term rewards\cite{Shi2019}. However, simulating the behaviors of hundreds of millions of customers in a dynamic environment is more challenging. In this work, Virtual Taobao is constructed by generating both customers and interactions. As shown in \textbf{Figure 18}, the GAN-SD (Generative Adversarial Network for Simulating Diversity) method is proposed to simulate different customers, including their requests; and a Multi-agent Adversarial Imitation Learning (MAIL) method is proposed for interaction generation. MAIL simultaneously learns both customer and platform policies, following the idea of GAIL (Generative Adversarial Imitation Learning), using a generative adversarial framework. MAIL trains a discriminator to distinguish simulated interactions from real ones. The discrimination signal is used as a reward and fed back to the customer and platform policies to generate more realistic interactions. Once customer and interaction generation are complete, Virtual Taobao is established.

\begin{figure}[htbp]
\centering
\includegraphics[width=\linewidth]{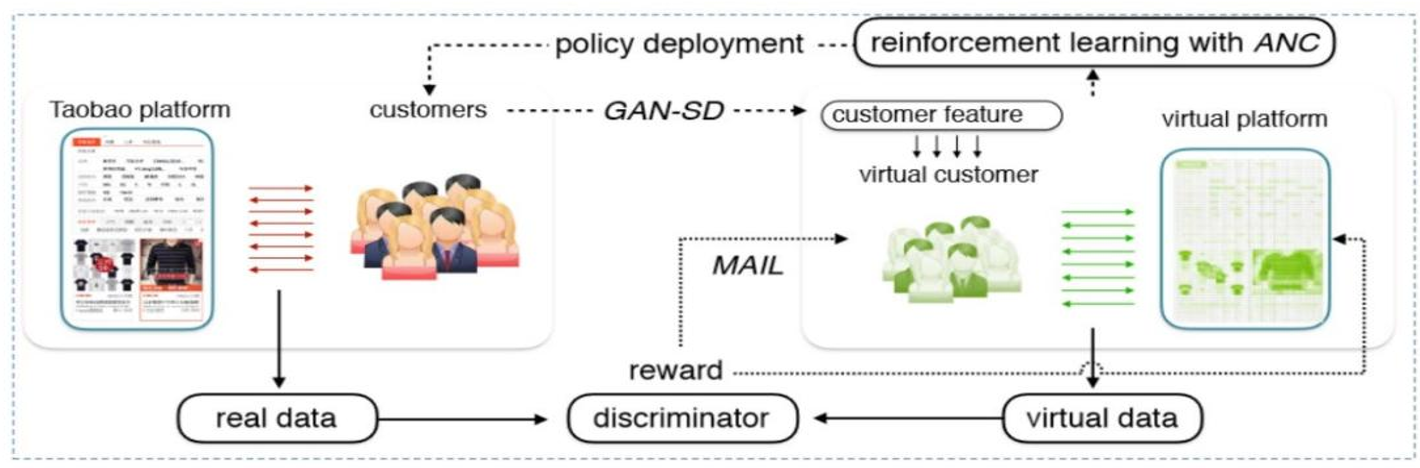}
\caption{Virtual Taobao Architecture Using Reinforcement Learning.}
\end{figure}

Experimental results show that Virtual Taobao successfully reconstructs attributes that closely resemble those of the real environment. Virtual Taobao can be used to train platform policies to maximize revenue. Compared to traditional supervised learning methods, the policies trained within Virtual Taobao achieved over a 2\% increase in revenue when deployed in the real environment.

\paragraph{Autonomous Driving}
Autonomous driving requires rigorous testing and validation of a vehicle’s ability to understand complex traffic scenarios and make driving decisions—this is one of the major challenges in the field of artificial intelligence. There are two main reasons for this: (1) Autonomous driving development demands vast amounts of data. According to an estimate by the RAND Corporation, a complete autonomous driving system must undergo at least 11 billion miles (approximately 17–18 billion kilometers) of validation before reaching mass production readiness. (2) Extreme scenario testing is rare and hard to reproduce. Constructing real-world scenarios such as blizzards, heavy rain, or typhoons is extremely challenging, and creating, replicating, and iterating such environments is costly—almost impossible to accomplish in practice.

Therefore, training and testing autonomous driving strategies in virtual environments has become a feasible technical solution. It is controllable, repeatable, and safe. To ensure that simulation results in the virtual world closely match those in the real world, three levels of fidelity must be achieved: (1) Geometric fidelity, which requires 3D scene simulation and sensor simulation, ensuring that both the environment and vehicle conditions mirror the real world. (2) Logical fidelity, which involves simulating the decision-making and planning processes of the test vehicle in the virtual world. (3) Physical fidelity, which requires simulation of the vehicle's control behavior and the outcomes of its dynamics. Additionally, the simulation platform must support high concurrency, allowing for permutations and combinations of vehicle responses across all scenarios. Currently, parallel learning methods are increasingly applied in virtual scenario generation and intelligent testing of autonomous vehicles\cite{Dugundji2008, DiMaggio2011, Fountain2014, Mare2011}.

\begin{figure}[htbp]
\centering
\includegraphics[width=\linewidth]{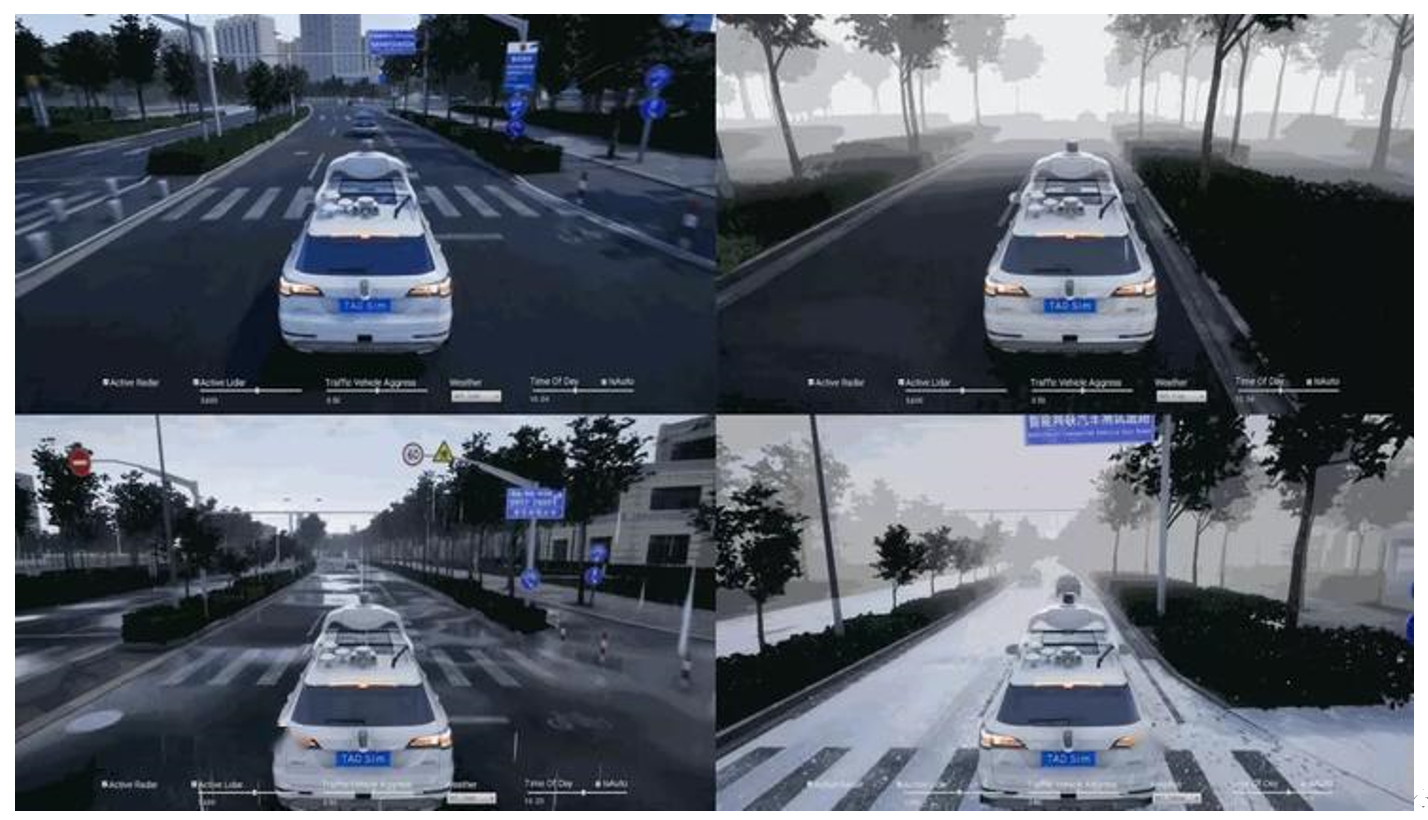}
\caption{Tencent TAD Sim Simulation System Scenario Demonstration.}
\end{figure}

As shown in \textbf{Figure 19}, Tencent has built a cloud-based virtual city that runs in parallel with the real physical world, leveraging its simulation platform, high-definition mapping platform, and data cloud platform. Urban simulation encompasses not only static environmental information but also dynamic data such as traffic and pedestrian flow. It can also integrate virtual elements such as traffic flow information. This supports both the development and safety validation of autonomous driving, while also contributing to the construction of smart cities and intelligent transportation systems. To improve the utilization of road data and enrich testing scenarios, technologies similar to Agent AI can be employed. By training traffic flow AI with large volumes of road-sensing data, highly realistic and interactive traffic scenarios can be generated for closed-loop simulation, enhancing testing efficiency and reducing data collection costs. For example, when an autonomous vehicle under test attempts to overtake, Agent AI can control non-player character (NPC) vehicles to perform realistic avoidance or other strategic behaviors consistent with real-world dynamics.

\section{Summary}
The first part of this paper addresses the challenges of quantification, analysis, and regulation in social-ecological system governance. From the perspective of complexity science, it proposes a technical pathway that integrates computational experimentation with real-world social scenarios. This pathway includes three key component models: agent model, environment model and rule model. Together, these components aim to provide technical support for the building of artificial society similar to real world. Finally, some classic cases of social simulation are given, covering three types: thought experiments, mechanism exploration and parallel optimization.

\section*{Acknowledgment}
This work has been supported in part by the National Natural Science Foundation of China (No.62472306, No.62441221,No.62206116), Tianjin University's  2024 Special Project on Disciplinary Development (No.XKJS-2024-5-9), Tian jin University Talent Innovation Reward Program for Lit erature \& Science Graduate Student (C1-2022-010), and Henan Province Key Research and Development Program (No.251111210500).

\bibliographystyle{IEEEtran}
\bibliography{2025.9.21}

\end{document}